\documentclass{article}
\usepackage{setspace}

\usepackage{epsfig}
\usepackage{graphics}
\usepackage{amssymb}
\usepackage{latexsym}
\usepackage{amsmath}
\begin{document}
\begin{center}

{\LARGE Numerical simulations of impacts involving porous bodies: \\I. Implementing sub-resolution porosity in a 3D SPH Hydrocode\\} 
\vspace {.5 cm}
{by\\}{
\vspace {.5 cm}
MARTIN JUTZI$^{1,2}$, WILLY BENZ$^{1}$,
PATRICK MICHEL$^{2}$\\}
\vspace {1. cm}
{$^1$ Physikalisches Institut, University of Bern, Sidlerstrasse 5, CH-3012
Bern, Switzerland \\}
{$^2$ University of Nice-Sophia Antipolis, Observatoire de la C\^ote d'Azur, UMR 6202 Cassiop\'ee/CNRS, B.P. 4229, 06304 Nice cedex 4,
France\\}
\vspace{.5 cm}
TEL: (+41) 31 631 4057 \\
FAX: (+41 31 631 4405) \\
E-MAIL: jutzi@space.unibe.ch\\
\end{center}
\vspace {1.5 cm}
\textbf{Length:\\} 60 manuscript pages\\2 Tables\\13 Figures
\newpage
\noindent\textbf{Running Title:\\} Impact simulations with porosity \\\\
\textbf{Corresponding author:\\}
Martin Jutzi\\
Physikalisches Institut \\
University of Bern \\
Sidlerstrasse 5 \\
CH-3012 Bern \\
Switzerland\\
TEL: (+41) 31 631 4057 \\
FAX: (+41) 31 631 4405\\
E-MAIL: jutzi@space.unibe.ch \\

\newpage

\begin{center}
{\LARGE Abstract\\}
\end{center}

In this paper, we extend our Smooth Particle Hydrodynamics (SPH) impact code to
include the effect of porosity at a sub-resolution scale by adapting the so-called $P-alpha$ model. 
Many small bodies in the different populations of asteroids and comets are believed to contain a high degree of porosity and 
the determination of both their collisional evolution and the outcome of their disruption requires that the effect of porosity is taken into account in the computation of those processes. Here, we present our model and show how porosity 
interfaces with the elastic-perfectly plastic material description and the brittle fracture model generally 
used to simulate the fragmentation of non-porous rocky bodies. We investigate various compaction models and discuss their suitability to simulate the compaction of (highly) porous material. Then, we perform simple test cases where we compare results of the simulations to the theoretical solutions. We also present a Deep Impact-like simulation to show the effect of porosity on the outcome of an impact. Detailed validation tests will be 
presented in a next paper by comparison with high-velocity laboratory experiments on porous materials (Jutzi et al., in preparation). 
Once validated at small scales, our new impact code can then be used at larger scales to study impacts and 
collisions involving brittle solids including porosity, such as the parent bodies of C-type asteroid families or 
cometary materials, both in the strength- and in the gravity-dominated regime.  

\newpage

\section{Introduction}

The collisional process plays a key role in all stages of a planetary system history, from the phase of 
planetary formation through collisional accretion to the late phases where the populations of small bodies 
evolve collisionally in a disruptive way. So far, these problems have been addressed using models appropriate 
to solid bodies represented as brittle rocky materials in which porosity was neglected or modeled at a macroscopic (i.e. resolved) level. Numerical codes, called 
{\it hydrocodes} (see, e.g., Benz and Asphaug, 1994) have been developed to compute the impact induced fragmentation of 
such solid bodies by solving the elastic-plastic conservation equations 
associated with a model of brittle failure to account for the fracture of the solid material. This already allowed to improve our understanding of the impact response of small bodies such as basalt-like 
material and asteroids belonging to the taxonomic class S, supposed to be composed of material with negligible porosity. However, several evidence point to the presence of a high degree of porosity in some 
small body classes. Asteroids belonging to the C taxonomic class are now believed to be highly porous, as 
indicated by the low bulk density ($\approx 1.3$ g$/$cm$^3$) estimated for some of them, such as the asteroid 
253 Mathilde encountered by the NEAR Shoemaker spacecraft (Yeomans et al., 1997), and 
as inferred from meteorite analysis (Britt et al., 2006). Small body populations 
evolving at larger distances from the Sun, i.e. the Jupiter-Family Comets, the Kuiper Belt Objects, and the 
other classes of comets contain also a high level of porosity, as indicated by the estimated low bulk 
densities (below $1$ g$/$cm$^3$, e.g., Rickman, 1998) and the analysis of interplanetary dust particles collected 
on Earth.
 
In parallel, the dissipative properties of porous media are invoked more and
more as playing an important role in the formation of early planetesimals (e.g., Wurm
et al., 2005). Hence, porosity emerges slowly as playing a major role from the time of the
formation of the planets to the collisional evolution of the present day Solar System.

Despite the growing focus on porosity, our ability to model its effect on the outcome of impacts and collisions remains limited. To address the question of modeling porosity in this context it is necessary to first define its scale in comparison with the other relevant dimensions involved in the problem such as the size of the projectile and/or the crater, etc.  In particular we define microscopic porosity as a type of porosity characterized by pores sufficiently small that one can reasonably assume that they are distributed uniformly and isotropically over these relevant scales. Macroscopic porosity on the other hand would be characterized by pores with sizes such that the medium can no longer be assumed to have homogeneous and isotropic characteristics over the scales of interest. In this case, pores have to be modeled explicitly and the current hydrocodes developed for the modeling of non-porous brittle solids  can still be used. The presence of these large macroscopic voids will only affect the transfer efficiency and the geometry of the shock wave resulting from the impact which can be computed using this existing software.
On the other hand, a body containing microporosity (i.e. porosity on a scale much smaller than can be numerically resolved) can be crushable: cratering on
a microporous asteroid might be an event involving compaction rather than ejection (Housen et al., 1999). In an impact in microporous material, a part of kinetic energy is dissipated by compaction which leads to less ejection and lower velocities of the ejected material. These effects cannot be reproduced using hydrocodes developed for the modeling of non-porous solids. Therefore, a model is needed which takes pore compaction into account.

Sirono (2004) proposed a model which can be used to study low velocity collisions of porous aggregates. Using this model, the author showed that the energy dissipation by compaction can lead to the sticking of dust aggregates. However, this model is not appropriate for high velocity impacts because of the simplicity of the equation of state (by construction, the pressure only depends on density).

Recently, Wuennemann et al. (2006) proposed a so-called $\epsilon-alpha$ model suitable to model porous material and for the use 
in hydrocodes. In this paper, we propose an alternative approach based on the so-called $P-alpha$ model 
(Herrmann, 1969) which we found to be more appropriate to use with our numerical method (see Sec.~\ref{sec:altmod}). Our 3D Smooth Particle Hydrodynamics (SPH) code, originally developed by Benz and Asphaug (1994), includes a model of brittle failure of non-porous material and 
successfully reproduced impact experiments on non-porous basalt targets. Moreover, associated to  
the N-body code {\it pkdgrav} to account for the effect of gravity (see e.g., Richardson et al., 2000), 
it reproduced successfully for the first time the formation of S-type 
asteroid families resulting from the catastrophic disruption of non-porous parent bodies (see, e.g.,  
Michel et al., 2001; Michel et al. 2003). A first implementation of a porosity model has been made in this code, 
by Benz and Jutzi (2006) using a simplified version of the $P-alpha$ model (the density $\rho$ was used instead of the 
pressure $P$) but was found to be inappropriate for materials with a high degree of porosity (see Sec.~\ref{sec:highpor}). Here, we improve on this first work by implementing the full $P-alpha$ model which offers several advantages and has a more general application.

In the following, we begin by describing our porosity model and comparing it to alternative models. We discuss some (counter intuitive) effects which occur when dealing with highly porous material. In the case of a high porosity, the energy dissipation during compaction leads to a large thermal pressure which can even become large enough to cause a decrease of the matrix density (Sec.~\ref{sec:highpor}). We show that our model is suitable to model these effects.
In Sec.~\ref{sec:nummod}, we recall the equations used in our modeling of brittle solids 
and then we show how porosity is related to these equations (Sec.~\ref{sec:distsolid}). 
In Sec.~\ref{sec:tests}, simple test cases are presented: first, a 
simulation of a 1D shock wave in porous media is compared to the analytical solution; in the second test, we 
simulate the compression of porous pumice and measure the so-called crush-curve. As a first application, we perform a simulation of a Deep Impact-like impact and compare simulations with and without porosity model. We also compare different compaction models using a porous basalt target. Conclusions are then exposed in Sec.~\ref{sec:concl}.
A detailed comparison of numerical simulations with high-velocity impact experiments on porous material is presented in a forthcoming 
paper (Jutzi et al., in preparation). 
 
\section{Porosity model} \label{sec:porosity}

While porosity at large scales can be modeled explicitly by introducing macroscopic voids, porosity on a scale 
much smaller than the numerical resolution has to be modeled through a different approach. Our porosity 
model is based on the $P-alpha$ model originally proposed by Herrmann (1969) and later modified 
by Carroll and Holt (1972). The model provides a description of microscopic porosity with pore-sizes beneath the 
spatial resolution of our numerical scheme (sub-resolution porosity) and which is homogeneous 
and isotropic.

\subsection{$P-alpha$ model}
The basic idea underlying the $P-alpha$ model consists in separating the volume change in  a  porous 
material into two parts: the pore collapse on one hand and the compression of the material composing the matrix 
on the other hand. This separation can be achieved by introducing the so-called distention parameter $\alpha$ defined as

\begin{equation}
\alpha\equiv\frac{\rho_s}{\rho}
\label{defalpha}
\end{equation}
where $\rho$ is the bulk density of the porous material and $\rho_s$ is the density of the corresponding 
solid (matrix) material. Distention is related to porosity as $1-1/\alpha$. According to its 
definition, the distention varies in the range $\alpha_0>\alpha>1$, where $\alpha_0$ is the initial distention.

Using the distention parameter $\alpha$, the equation of state (EOS) can be written in the general form:
\begin{equation}
     P=P(\rho,E,\alpha)
     \label{geos}
\end{equation}
According to Caroll and Holt (1972), the EOS of a porous material can explicitly be written as:
\begin{equation}
     P=\frac{1}{\alpha} P_s(\rho_s,E_s) = \frac{1}{\alpha} P_s(\alpha
      \rho, E)
\label{meos}
\end{equation}
where $P_s(\rho_s,E_s)$ represents the EOS of the solid phase of the material (the matrix). A crucial
assumption in this model is that the pressure depends on the density of the matrix material. The pore space is modeled as empty voids 
and the internal energy $E$ is assumed to be the same in the porous and the solid material ($E=E_s$), which implies that the surface 
energy of the pores is neglected. The factor $1/\alpha$ in Eq.~(\ref{meos}) was introduced by Carroll and Holt (1972) to take into account that the volume average of the stress in the matrix is given by 
\begin{equation}
 P_s=\alpha P
\end{equation}
where $P$ is the applied pressure.

In the $P-alpha$ model, the distention is solely a function of the pressure $P$:
\begin{equation}
      \alpha=\alpha(P)
      \label{eq:alphap}
\end{equation}
where $P = P(\rho,E,\alpha)$. The relation between distention and pressure is often divided in an elastic regime   
($P<P_e$) and a plastic regime ($P>P_e$), where $P_e$ is the pressure at which the transition between the two 
regimes occurs. In the elastic regime, the change of $\alpha$ with pressure is reversible. According to Herrmann (1969), the relation between distention and pressure can be defined as  
\begin{equation}\label{dadpe}
       [\frac{d\alpha}{dP}]_{elastic} =
\frac{\alpha^2}{K_0} [1-(\frac{1}{h(\alpha)^2})]
\end{equation}
where $K_0=c_0^2\rho_0$ and
\begin{equation}
       h(\alpha)=1+(\alpha-1) \frac{c_e-c_0}{c_0(\alpha_e-1)}
\end{equation}
where $\alpha_e=\alpha(P=P_e)$. This equation follows from the assumption that the elastic wave velocity 
changes linearly from the initial value $c_e$ to the bulk sound speed $c_0$ in the solid state. 
Using Eq.~(\ref{dadpe}) leads to a small change of the distention (from $\alpha_0$ to $\alpha_e$) in the elastic regime and an elastic wave velocity which is smaller than in the case of a constant $\alpha$. Since the distention changes only very little, it is often assumed that $d\alpha/dP=0$ in the elastic regime and therefore $c_e=c_0$.

In the plastic regime, the following quadratic form is often used to define the function $\alpha=\alpha(P)$:
\begin{equation}
    \alpha =  1 + (\alpha_e-1)
     \frac{(P_s-P)^2}{(P_s-P_e)^2}.
     \label{alphaqudratic}
\end{equation}
where $P_e$ and $P_s$ are constant. 

This is obviously a very simple model, but it is appropriate enough for many applications. A more 
realistic relation can be obtained experimentally by means of a one dimensional static compression of a sample, 
during which the actual distention $\alpha_m$ is measured as 
a function of the applied pressure $P_m$. The resulting 
crush-curve $\alpha_m(P_m)$ then provides the required relation between distention and pressure for the 
material. In Sec.~\ref{sec:tests}, we present test simulations where we use such a measured relation to define 
$\alpha(P)$.

\subsection{Alternative models} \label{sec:altmod}

According to the $P-alpha$ model, distention is a function of pressure $\alpha=\alpha(P)$. However, distention can also be defined as a function of other state variables. Wuennemann et al. (2006) for example use the volumetric strain ($\epsilon-alpha$ model). Another 
possibility is to define distention directly as a function of density. We will refer to that approach as $\rho-alpha$ model. Note that in each model, the pressure is computed according to Eq.~(\ref{meos}).

\subsubsection{$\epsilon-alpha$ versus $\rho-alpha$ model}

According to Wuennemann et al. (2006), the volumetric strain can be expressed as
\begin{equation}
 \epsilon_v=\int_{V_0}^V \frac{V'}{dV'} = ln(V/V_0)
\end{equation}
where $V_0$ is the initial volume and $V$ the actual volume. Assuming that the volume of the matrix is kept constant ($V_s=V_{s0}$) volumetric strain and 
distention can be related as $\epsilon_v=ln(\alpha/\alpha_0$), which leads to the compaction function
\begin{equation}
\alpha=\alpha_0 e^{\epsilon_v}
\label{eq:ae}
\end{equation}
Another (very similar) way to define the evolution of the distention follows from its
definition: $\alpha=\rho_s/\rho$. If we again assume a constant matrix volume and therefore a constant matrix 
density ($\rho_s=\rho_{s0}$), we get the compaction function
\begin{equation}
\alpha=\frac{\rho_{s0}}{\rho}.
\label{eq:arho}
\end{equation}
This relation also follows from Eqs. (\ref{eq:ae}) by replacing $\epsilon_v$ by
\begin{equation}
 \epsilon_v=ln(\frac{V}{V_0})= ln(\frac{\rho_0}{\rho})
\label{eq:evrho}
\end{equation}
where $\rho_0=\rho_{s0}/\alpha_0$ is the initial density.

Both equations, (\ref{eq:ae}) and (\ref{eq:arho}), describe the fasted possible way to decrease distention as a function of volume (density) change under the assumption that the matrix density is constant. However, as Wuennemann et al. (2006)
point out, during the compression of a porous material, not only pore space is compacted, but 
also the matrix is slightly compressed leading to an increase of the pressure. Wuennemann et al.   
(2006) take this into account by introducing a parameter $\kappa$ to control the compaction rate:
\begin{equation}
\alpha=\alpha_0 e^{\kappa (\epsilon_v-\epsilon_e)}
\label{eq:aemod}
\end{equation}
where $\epsilon_e$ is the critical strain where the compaction starts.
Using Eq. (\ref{eq:evrho}) and defining $\epsilon_e \equiv ln(\rho_0/\rho_e)$, we can write this equation in terms of density
\begin{equation}
\alpha=\alpha_0 (\frac{\rho_e}{\rho})^\kappa.
\label{eq:arhomod}
\end{equation}
Wuennemann et al. (2006) also use a power-law compaction regime at a certain volumetric strain. In this way, they are able to 
reproduce experimental compaction data obtained by a uniaxial compression of a porous 
material. In a similar way, one could also modify Eq.~(\ref{eq:arho}) to obtain similar results. 

In both models, the time evolution of the distention parameter has a very simple form. In the $\epsilon-alpha$  
model, it is given by
\begin{equation}
 \dot\alpha=\frac{d\alpha}{d\epsilon}\dot\epsilon
 \label{dadededt}
\end{equation}
and for the $\rho-alpha$ model we get
\begin{equation}
 \dot\alpha=\frac{d\alpha}{d\rho}\dot\rho
 \label{dadrdrdt}
\end{equation}
The comparison above shows that the $\epsilon-alpha$ model and the $\rho-alpha$ model are very similar and the 
only difference is the parameter which is chosen to measure the actual volume or density, respectively.

\subsubsection{$\rho-alpha$ versus $P-alpha$ model}\label{sec:rapa}

In the $P-alpha$ model, the distention depends on the density via the pressure 
given by Eq.~(\ref{meos}) but contrary to the two models described above, it is also a function of the 
internal energy. However, for small initial porosities, the energy contribution to the pressure 
remains small as long as $\alpha>1$,  and can even be neglected in most cases. Therefore, the pressure can be 
approximated by $P\simeq P(\rho,\alpha)$ and consequently, we can 
transform the function $\alpha(P[\rho,\alpha])$ in $\alpha=\alpha(\rho)$. In this way, we can define a function $\alpha=\alpha(\rho)$ which, instead of mimicking the behavior of the $\epsilon-alpha$ model as in Eq.~(\ref{eq:arhomod}),
approximately corresponds to the relation $\alpha=\alpha(P)$, but neglects the thermal contribution to the pressure. Test simulations aimed 
at comparing $\alpha=\alpha(P)$ and $\alpha=\alpha(\rho)$ show that for low porosities this assumption is valid and the difference between the two models is rather small (see Sec.~\ref{sec:tests}).
However, for high porosities this assumption is not valid and we observed some problems using this
form of the $\rho-alpha$ model.

\subsection{Problems with high porosities}\label{sec:highpor}

\begin{figure}[h!]
\centering
\resizebox{10cm}{!}{\includegraphics{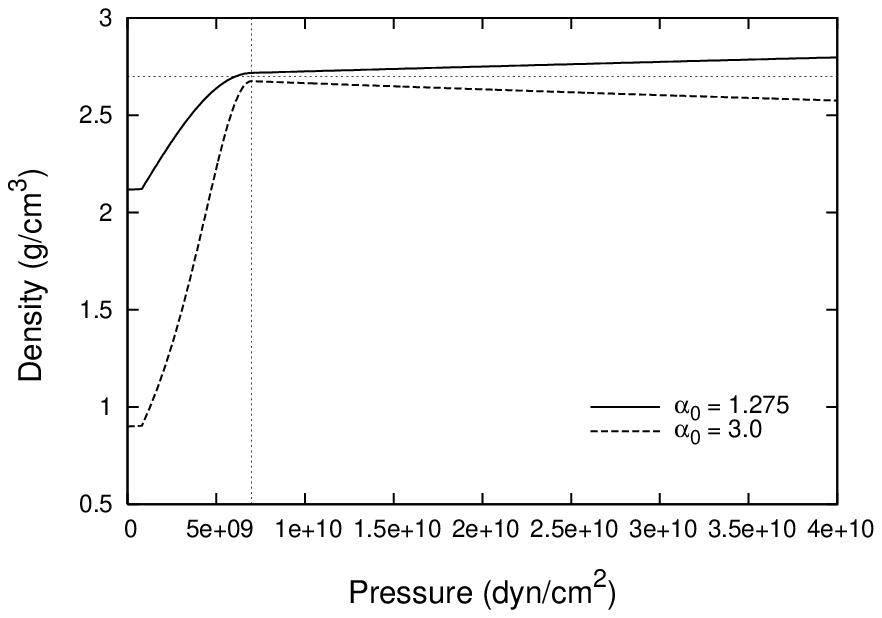} }
\vspace{1cm}
\resizebox{10cm}{!}{\includegraphics{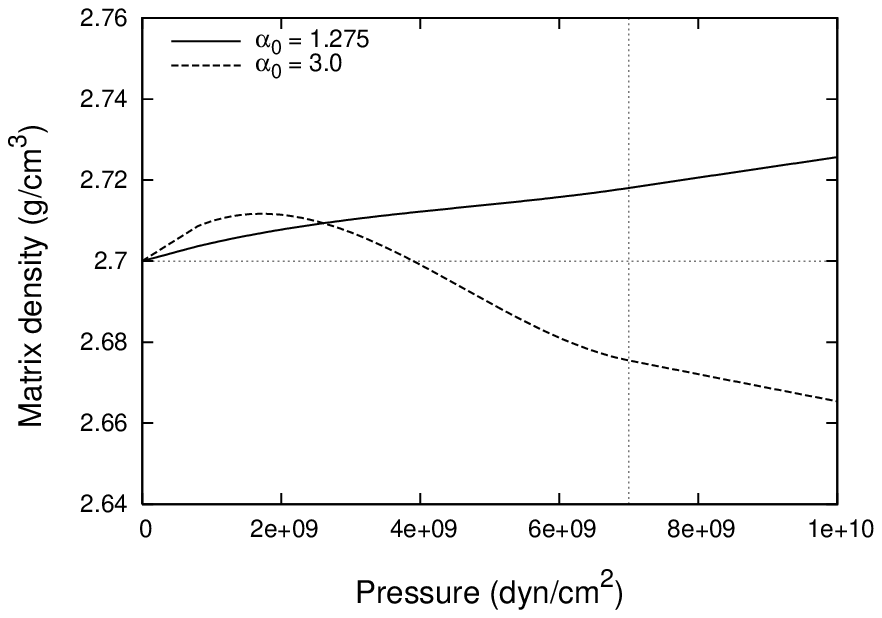} }
\caption{}
\label{fig:hugprho}
\end{figure}

As we described above, the energy contribution to the pressure is very small in the porous regime ($\alpha > 1$). However, 
this is only true for small initial porosities. In highly porous material, the thermal pressure can become 
large enough that the density in the compressed state ($\alpha=1$) is below the initial density of 
the matrix (Zel'dovich and Raizer, 1967), while in a less porous material the density at $\alpha=1$ would be 
at least as high as the initial density of the matrix. Then, when the compressed state is reached, 
the density does not increase further with increasing pressure but rather decreases, and the volume increases 
accordingly. This behavior can lead to an anomalous but nonetheless actual Hugoniot curve (see Fig.~\ref{fig:hugprho}, top). It is important to stress that this anomalous behavior of highly porous material is not only a theoretical concept but can actually be observed in experiments.

In the fully compressed state, by definition the material is only composed of the matrix. Since in this state the density of highly porous material is smaller than the initial value of the matrix density, this implies that the matrix density in the fully compacted state has decreased from its initial value. Such a decrease must occur even before this fully compressed state has been reached
(see Fig.~\ref{fig:hugprho}, bottom). 
As an important consequence, in highly porous materials the distention can be decreased faster (as a function of volume change) than in the case of a constant matrix density (Eqs.~(\ref{eq:ae}) and (\ref{eq:arho})). This behavior occurs independently of the actual functional form of the distention. It is actually caused by the huge difference between the density of the porous material and the matrix density. 

It is important to note that even in highly porous material, the change of the matrix density during the compaction is very small compared to the change of the bulk density (see Fig.~\ref{fig:hugprho}). Nevertheless, it has a great effect since according to Eq.~(\ref{meos}), the matrix density is used to compute the pressure.  

To correctly simulate the compaction of highly porous materials, therefore, the compaction
functions must allow the density of the matrix material to decrease. For the $\rho - alpha$ or $\epsilon - alpha$
models this implies that, above some threshold pressure, the distension must decrease faster (as a
function of volume change) than defined by Eq. (\ref{eq:ae}) or (\ref{eq:arho}), which follow from the assumption
of a constant matrix density. This could be achieved by a modification of Eq. (\ref{eq:aemod}) or (\ref{eq:arhomod}). On the
other hand, defining distension as a function of pressure offers a simple prescription for
compaction that allows for matrix expansion. The $P - alpha$ model, without modification, can therefore
simulate the compaction of highly porous materials given suitable model parameters.

\begin{figure}[h!]
\centering
\resizebox{10cm}{!}{\includegraphics{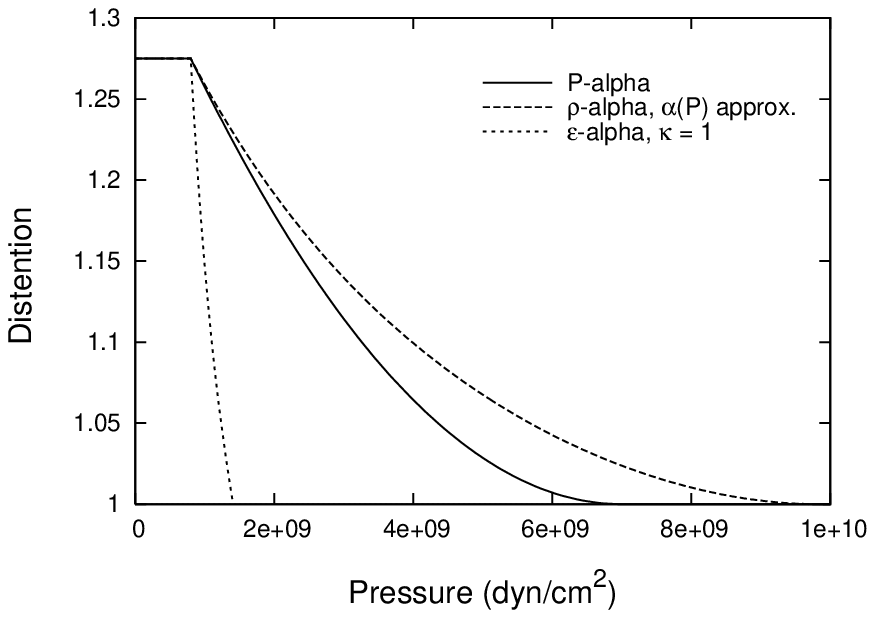} }
\vspace{1cm}
\resizebox{10cm}{!}{\includegraphics{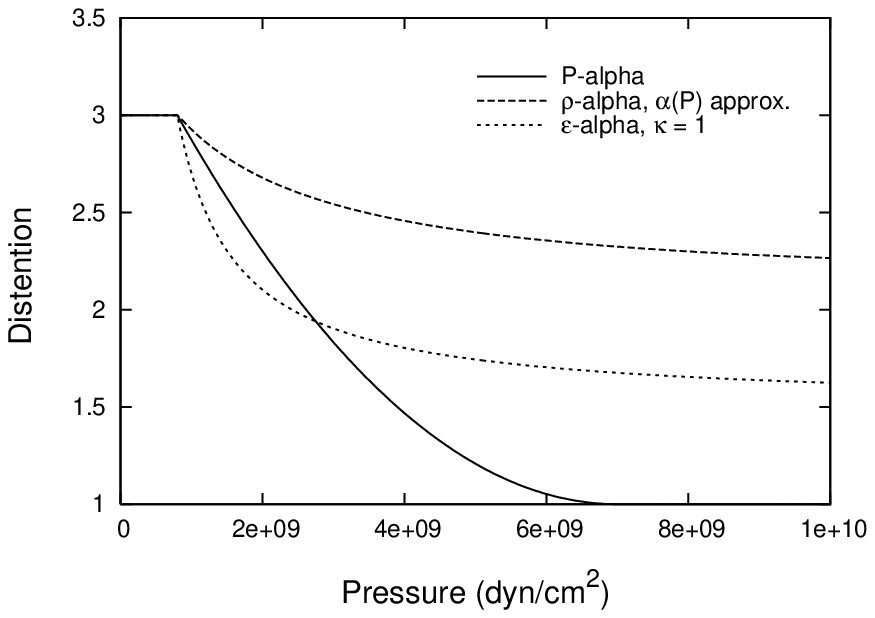} }
\caption{}
\label{fig:hugcomp}
\end{figure}

For illustration purposes, we show the pressure-distention relation in a unidimensional shockwave
under three different assumptions of compaction behavior, and for two initial porosities (Fig.~\ref{fig:hugcomp}). In
the first case we use the $P-alpha$ model which assumes a quadratic function (Eq.~\ref{alphaqudratic}) for $\alpha=\alpha(P)$ with
$P_e$=8$\times$10$^8$ dyn/cm$^2$ and $P_s$=7$\times$10$^9$ dyn$/$ cm$^2$, and that $\alpha_e = 0$ in the elastic regime. The values of $P_e$ and $P_s$ approximately correspond to the ones used by Herrmann (1969) to study a (low porosity) compaction wave in porous aluminium. For illustrative purposes, we assume that $P_e$ and $P_s$ do not depend on the initial porosity.
In the second case
we assume a constant matrix density during compaction by using Eq. (\ref{eq:ae}), with $\kappa = 1$. The density
$\rho_e$ is chosen so that the compaction starts at $P = P_e$. As a third case, we also show the results
obtained using a function $\alpha=\alpha(\rho)$ which follows from the assumption that the thermal contribution to
the pressure can be neglected and that the transformation of $P \simeq P(\rho,\alpha)$ holds, as described in the previous
section.

Since we only want to study the compaction behavior which takes place at moderate pressures ($P \lesssim P_s$) we use a simplified version of the Tillotson equation (Tillotson 1962, Melosh 1989) for this illustration. This simplified EOS provides a reasonable approximation of the full Tillotson equation in the considered regime. Furthermore, it allows us to examine the sensitivity of the pressure-distention relations on the equation of state (EOS) parameters. The following equation is used:
\begin{equation}\label{eostpcmod}
     P = c \rho E + A\mu
\end{equation}
where $c=a+b$, $\mu=\eta-1$ and $\eta=\rho / \rho_0$ and $A$, $a$ and $b$ are the usual Tillotson parameters. 
The assumptions made to obtain this equation are discussed in the appendix.

The pressure-distention relation is finally computed using Eq.~(\ref{meos}), the simplified EOS (Eq.~\ref{eostpcmod}) and the Hugoniot equation
\begin{equation}\label{henergycons}
E-E_0=(P+P_0)(V_0-V)/2
\end{equation}
where $V_0=1/\rho_0$ and $V=1/\rho$ and $P_0$ = 0 and $E_0$ = 0 are used as initial values.
We further use $c$ = 2 and $A$ = 7.5$\times$10$^{11}$ dyn/cm$^2$ (these values approximately correspond to the usual aluminium parameters) for the illustration (Fig.~\ref{fig:hugcomp}).

We have to point out the the following considerations are only valid in a moderate pressure regime where the assumptions leading to the simplified EOS are reasonable. 

We show two cases with a different initial distention, $\alpha_0$ = 1.275 and $\alpha_0$ =
3.0. In the low porosity case, full compaction is reached with all three models at similar pressures.
Of course, the curve obtained by using the $P - alpha$ model corresponds to the quadratic relation
used to define $\alpha(P)$. The curve denoted by $\rho - alpha$ shows the results obtained using the
approximation $P(\rho,\alpha,E)\simeq P(\rho,\alpha)$ followed by the transformation $\alpha(P) = \alpha(P[\rho,\alpha]) \rightarrow \alpha = \alpha(\rho)$, which
assumes that the thermal contribution to the pressure can be neglected. As expected for low
porosities, it has a similar shape as the $P - alpha$ curve because in this case the thermal component
of the pressure is small. The difference in pressure between the two curves is equivalent to the
thermal contribution to the pressure at a given level of compaction. Note also that the $P - alpha$
curve is less steep than the constant matrix-density curve ($\epsilon - alpha$ model curve with $\kappa = 1$) because
in this case the $P - alpha$ relationship implies some compression of the matrix material during
compaction.

In the high porosity case, on the other hand, full compaction is achieved in the considered regime only
with the $P - alpha$ model. The curves of the other two models do not reach $\alpha$ = 1 until a much higher
pressure is reached. Even the assumption of constant matrix density does not lead to full
compaction. 
As described above, the difficulty in reaching complete compaction is caused by the fact
that the matrix density can actually decrease with increasing pressure (due to the large thermal component) for highly porous materials. 
In the case of the constant matrix density, it can be shown that the minimal value that the distention can reach is given by
\begin{equation}\label{eq:aminc}
 \alpha_{min}= \lim \limits_{P \to \infty} \alpha(\epsilon[P]) = \alpha_0\frac{\rho_e}{\rho_0}\frac{c}{c+2} \simeq \alpha_0\frac{c}{c+2}
\end{equation}
We have to point out that this result follows using the simplified EOS (Eq.~\ref{eostpcmod}) which is of course not valid for infinite pressures. Using the full Tillotson equation of state, the contribution of the term $\mu^2 B$ and the reduction of the effective value of $c$ at very high pressures (much higher than 10$^{10}$ dyn/cm$2$) can still lead to full compaction.

\begin{figure}[h!]
\centering
\resizebox{10cm}{!}{\includegraphics{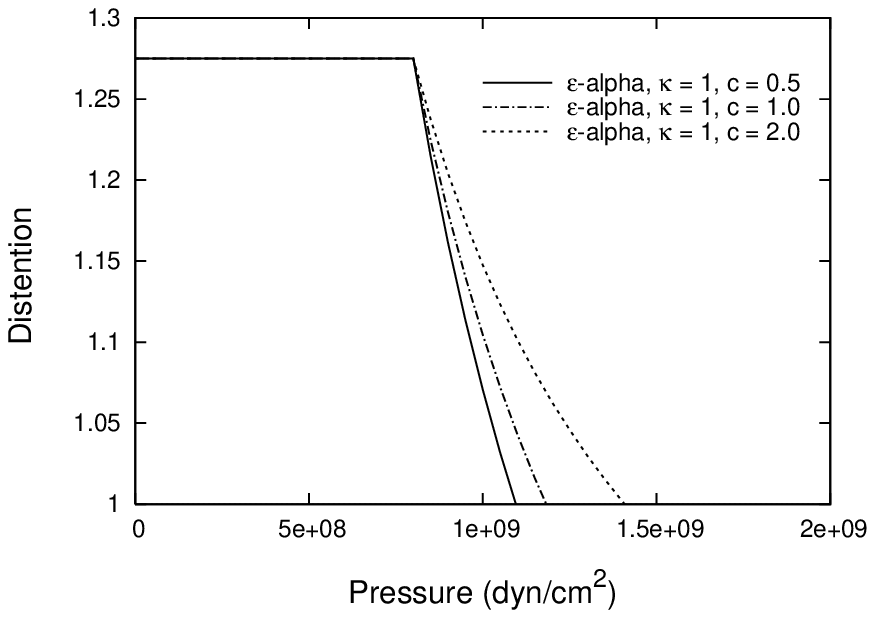} }
\vspace{1cm}
\resizebox{10cm}{!}{\includegraphics{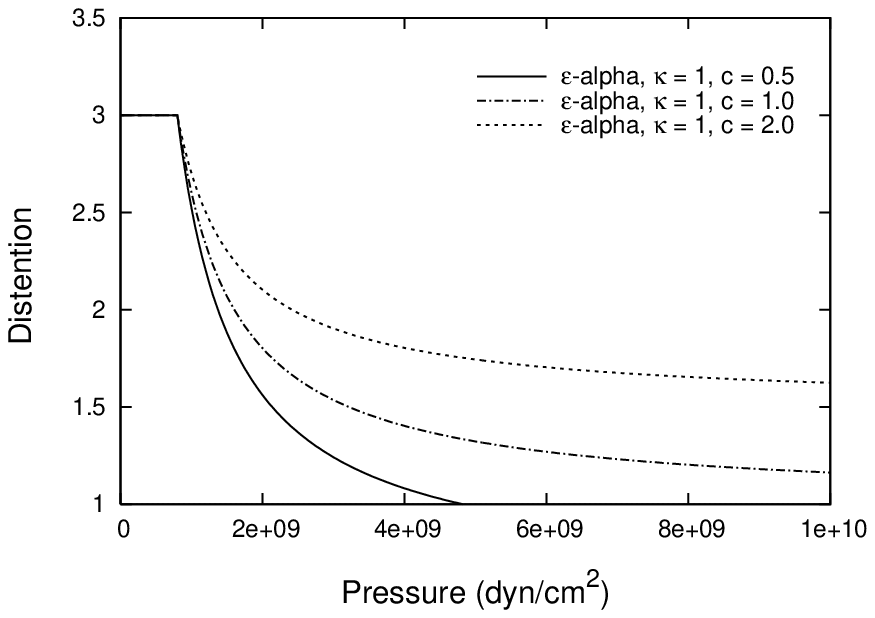} }
\caption{}
\label{fig:hugcompevc}
\end{figure}

As Eq.~(\ref{eq:aminc}) shows, the parameter $c$ determines whether or not full compaction is possible at moderate pressures. For illustration, we once again show the curves which follow from the constant matrix-density assumption but now for different values of $c$ (again for two cases with $\alpha_0$ = 1.275 and $\alpha_0$ = 3.0). For small porosities, changing $c$ from 0.5 to 2 only slightly changes the slope of the curves (see Fig.~\ref{fig:hugcompevc}, top). On the other hand, in the high porosity case, the parameter $c$ has a great influence on the crushing behavior (see Fig.~\ref{fig:hugcompevc}, bottom). For $c$ = 0.5, full compaction is reached at $P\simeq$ 5$\times$10$^9$ dyn/cm$^2$. However, for $c$ = 1 and $c$ = 2 we get an $\alpha_{min}$ of 1.0 and 1.5, respectively. The corresponding curves asymptotically approach these values. Again, using the full Tillotson equation, the  behavior would be different for very high pressures and full compaction  would be achieved even for $c$ = 2. Nevertheless, we think that the
estimation of $\alpha_{min}$ (Eq.~\ref{eq:aminc}) can be used at least as a first guess of value of a critical distention
\begin{equation}\label{eq:acrit}
 \alpha_{0crit} \simeq \frac{c+2}{c}
\end{equation}
which follows from $\alpha_{min} = 1$.
For an initial distention higher than this value ($\alpha_0 > \alpha_{0crit}$), anomalous effects can occur and models which do not allow matrix expansion fail to reach full compaction at moderate pressures. Obviously, this procedure to determine $\alpha_{0crit}$ only works for EOS where a parameter $c$ can be identified.  In any case, the results indicate that using models where the distention is decoupled from pressure ($\epsilon - alpha$ and $\rho - alpha$ model), the compaction behavior can strongly depend on the EOS parameters which relate energy and pressure. On the other hand, using a model where the distention is a direct function of the pressure ($P - alpha$ model), the compaction behavior is not sensitive to these parameters.

In Sec.~\ref{sec:deepimp}, results of impact simulations using the $P - alpha$, $\rho - alpha$ and $\epsilon - alpha$ model are compared. 

\subsection{Our actual model}

Although the use of the $\rho-alpha$ model (as an analogue to the $\epsilon-alpha$ model) has some advantages, 
especially the simple form of the time evolution $\dot\alpha=\frac{d\alpha}{d\rho}\dot\rho$, we found that for 
our numerical scheme,  the most appropriate variable to define a functional form of the distention is pressure: 
$\alpha=\alpha(P)$. The following reasons support this choice:

\begin{itemize}

\item The $P-alpha$ model can be used \textit{without modification} to simulate the compaction of highly porous material.

 \item The relation between distention and pressure, which is used as an input in our model, can directly be obtained from the 
experimental crush-curve for the material considered.

\item We found that using our numerical scheme, no iteration is needed to implicitly solve for pressure and distention.

\end{itemize}
For the actual form of the relation between distention and pressure we either use a quadratic relation 
(\ref{alphaqudratic}) or, if available, we directly use the experimentally measured crush-curve to define 
the function $\alpha(P)$. By definition, the pressure distention relation ($\alpha(P)$) we actually use in our code does not depend on the strain rate. Using  a crush-curve which was measured under quasi static conditions therefore assumes that the same curve would be obtained at high strain rates. Experiments (Nakamura et al., in preparation) show that there actually is a (small) strain rate dependence of the crushcurve. However, we found (Jutzi et al., in preparation) that the results of impact simulations (i.e., the fragment mass distribution) do not strongly depend on the exact shape of the  $P-alpha$ relation. Therefore, we think that it is not problematic to use a low strain rate crush-curve as input in our code to simulate high strain rate events.

The following function allows a good fit to a wide 
range of experimental crush-curves:
\begin{equation}
    \alpha(P) =  \begin{cases}
(\alpha_e-\alpha_t) \frac{(P_t-P)^{n1}}{(P_t-P_e)^{n1}} + (\alpha_t-1) \frac{(P_s-P)^{n2}}{(P_s-P_e)^{n2}} 
+ 1 & \text{if } P_e<P<P_t \\
(\alpha_t-1) \frac{(P_s-P)^{n2}}{(P_s-P_e)^{n2}} + 1 & \text{if } P_t<P<P_s \\
1 & \text{if } P_s < P \\
\end{cases}
     \label{alpha2reg}
\end{equation}
where $P_s,P_e$ and $\alpha_e$ have the same meaning as in Eq.~(\ref{alphaqudratic}), and $P_e<P_t<P_s$ 
and $1<\alpha_t<\alpha_0$ are parameters indicating a transition pressure and distention, respectively. The function (\ref{alpha2reg})  and its first derivative are smooth by definition, which allows the existence of 
two regimes of $\alpha(P)$, each with a individual slope ($n_1$ and $n_2$). 

In the elastic region ($P < P_e$) we either use Eq.~(\ref{dadpe}) to define $[d\alpha/dP]_{elastic}$ or, to simplify matters, we assume that the distention is constant, i.e. $\alpha_e=\alpha_0$. For all simulations presented in this paper (except the one in Sec.~\ref{sec:1dwave}), we use $[d\alpha/dP]_{elastic}=0$ in the elastic regime.

The derivative $d\alpha/dP$ which is used to compute the time evolution of the distention is now given by
\begin{equation}\label{dadpdef}
 d\alpha/dP =
\begin{cases}
[d\alpha/dP]_{elastic} & \text{if } P<P_e\\
[d\alpha/dP]_{plastic}& \text{otherwise}
\end{cases}
\end{equation}
where $[d\alpha/dP]_{plastic}$ follows from Eq.~(\ref{alpha2reg}).
We further assume that unloading (from a partially compacted) state is elastic. Consequently, we define
\begin{equation}\label{qconstraint}
 d\alpha/dP =
\begin{cases}
d\alpha/dP & \text{if } dP>0\\
[d\alpha/dP]_{elastic} & \text{otherwise.}
\end{cases}
\end{equation}
As discussed above, $[d\alpha/dP]_{elastic}$ is either computed using Eq.~(\ref{dadpe}) or assumed to be zero.

The time evolution of the distention parameter can be written as
\begin{equation}
 \dot\alpha=\frac{d\alpha}{dP}\frac{dP}{dt}
\end{equation}
Using Eq.~(\ref{meos}) we finally get
\begin{equation}
\dot \alpha(t)= \frac{\dot E \left(\frac{\partial P_s}{\partial E_s}\right) + \alpha \dot \rho 
\left(\frac{\partial P_s}{\partial \rho_s}\right)}{\alpha + \frac{d\alpha}{dP} \left[P - \rho 
\left(\frac{\partial P_s}{\partial \rho_s}\right)\right]}\cdot \frac{d\alpha}{dP}
\label{dadt}
\end{equation}

The equations (\ref{meos}) and (\ref{alpha2reg} - \ref{qconstraint}) define the constitutive equation which describes the compaction behavior of a porous material. In the original work of Herrmann (1969) and Carroll and Holt (1972), the $P-alpha$ model was intended to be a first order theory in which shear strength effects are considered secondary and consequently, the stress tensor was assumed to be diagonal. In this work, we use the full stress tensor and therefore, we extend the original model with a relation between the distention and the deviatoric stress tensor (Sec.~\ref{sec:diststrength} and \ref{pord}).

\section{Model equations and implementation}\label{sec:nummod}

Our numerical technique is based on the Lagrangian Smooth Particle Hydrodynamic (SPH) method. Since 
the basic method has already been described in many papers (see for examples reviews by Benz, 1990; Monaghan, 1992) 
we refer the interested reader to these earlier papers.

The standard gas dynamics SPH approach was extended (see for example Libersky and Petschek, 1991) to include 
an elastic-perfectly plastic material description and a fracture model based on the one of Grady and Kipp (1980) 
in order to model the behavior of brittle solids (Benz and Asphaug, 1994). As our porosity model
interfaces with this material description, we begin with a short review of this previous approach. Note that the following equations describe non-porous material. Porosity is introduced in Sec.~\ref{sec:distsolid}.

\subsection{Elastic perfectly plastic strength model}

The equations to be solved are the well-known conservation equations of elasto-dynamics; they can be found in 
most standard textbooks. The mass conservation can be written as:
\begin{equation}
   \frac{d\rho^{\kappa}}{dt}+\rho\frac{\partial v^{\kappa\lambda}}{\partial 
   x^   {\lambda}}=0
   \label{eq:massconv}
\end{equation}
where $d/dt$ is the Lagrangian time derivative, $\rho$ the density, $v$ the velocity and $x$ the position. 
The conservation of momentum has the following form:
\begin{equation}
   \frac{dv^{\kappa}}{dt}=\frac{1}{\rho}\frac{\partial\sigma^{\kappa\lambda}}
   {\partial x^{\lambda}}
\end{equation}
where $\sigma^{\kappa\lambda}$ is the stress tensor given by
\begin{equation}
\sigma^{\kappa\lambda}=S^{\kappa\lambda}-P\delta^{\kappa\lambda}
\end{equation}
where $P$ is the hydrostatic pressure, $\delta^{\kappa\lambda}$ is the Kroneker symbol 
and $S^{\kappa\lambda}$ is the (traceless) deviatoric stress tensor.
Finally, the conservation of energy is given by the equation
\begin{equation}
\frac{dE}{dt}=-\frac{P}{\rho}\frac{\partial}{\partial x^{\kappa}}v^{\kappa}+\frac{1}{\rho} S^{\kappa\lambda}
\dot\epsilon^{\kappa\lambda}
\end{equation}
where $\dot\epsilon$ is the strain rate tensor given by
\begin{equation}
\dot\epsilon^{\kappa\lambda}=\frac{1}{2}\left(\frac{\partial v^{\kappa}}{\partial x^{\lambda}}+\frac{\partial 
v^{\lambda}}{\partial x^{\kappa}}\right).
\end{equation}
In order to specify the time evolution of the deviatoric stress tensor $S^{\kappa\lambda}$ we adopt Hooke's 
law and define the time evolution of the deviatoric stress tensor as:
\begin{equation}\label{eq:ds}
\frac{dS^{\kappa\lambda}}{dt}=2\mu\left(\dot\epsilon^{\kappa\lambda}-\frac{1}{3}\delta^{\kappa\lambda}
\dot\epsilon^{\nu\nu}\right)+S^{\kappa\lambda}\Omega^{\lambda\nu}+S^{\lambda\nu}\Omega^{\kappa\nu}
\end{equation}
where $\mu$ is the shear modulus, and $\Omega$ is the rotation rate tensor:
\begin{equation}
\Omega^{\kappa\lambda}=\frac{1}{2}\left(\frac{\partial v^{\kappa}}{\partial x^{\lambda}}-\frac{\partial 
v^{\lambda}}{\partial x^{\kappa}}\right).
\end{equation}
Finally, plasticity is treated using the von Mises yielding criterion.

In order to solve this set of equations, an equation of state has to be specified which relates density, 
energy and pressure:

\begin{equation}
P=P(\rho,E)
\end{equation}
For the simulations presented in this paper we use the so-called Tillotson equation of state (e.g., Tillotson 1962, Melosh 1989).

\subsection{Fracture}

Brittle materials cannot be modeled using elasticity and plasticity alone because these materials fail under 
tension or shear stress. To take this behavior into account, we use a fracture model introduced by Grady and Kipp  
(1980) and based on the presence 
of incipient flaws in the material and on crack propagation under increasing strain. This model 
has been introduced using an explicit distribution of incipient flaws by Benz and Asphaug (1994, 1995) in their 
SPH code. In this model it is assumed that the number density of active flaws at strain $\epsilon$ is 
given by a Weibull distribution (Weibull, 1939)
\begin{equation}
n(\epsilon)=k\epsilon^m
\label{wbd}
\end{equation}
where $k$ and $m$ are the material dependent Weibull parameters. 
When the local tensile strain has reached 
the activation threshold of a flaw, a crack is allowed to grow at a constant velocity 
$c_g$ which is some fraction of the local sound speed. The half length of a growing crack is therefore
\begin{equation}
a=c_g(t-t')
\label{crackgr}
\end{equation}
where $t'$ is the crack activation time.

Crack growth leads to a release of local stresses. To model this behavior, we follow Benz and Asphaug (1994, 1995) 
and introduce a state variable $D$ (for damage) which expresses the reduction in strength under tensile 
loading:
\begin{equation}\label{eq:sd}
\sigma_D=\sigma(1-D)
\end{equation}
where $\sigma$ is the elastic stress in the absence of damage and $\sigma_D$ is the damage-relieved stress. The 
state variable $D$ is defined locally as the fractional volume that is relieved of stress by local growing cracks
\begin{equation}
D=\frac{\frac{4}{3}\pi a^3}{V}
\label{defD}
\end{equation}
where $V=4/3\pi R^3_s$ is the volume in which a crack of half - length $R_s$ is growing. 
Using Eqs.~(\ref{crackgr}) and (\ref{defD}) we get the following equation for the damage growth
\begin{equation}\label{Dgr}
\frac{dD^{1/3}}{dt}=\frac{c_g}{R_s}
\end{equation}
Damage accumulates at a rate given by Eq.~(\ref{Dgr}) when the local tensile strain $\epsilon_i$ reached the activation threshold of a flaw. Note that $\epsilon_i$ is obtained from the maximum tensile stress $\sigma^t_i$ after a principal axis transformation
\begin{equation}
 \epsilon_i=\frac{\sigma^t_i}{(1-D_i)E}
\label{eq:epsi}
\end{equation}
where $D_i$ is the local value of the damage and $E$ is the Young modulus. 

\section{Interfacing porosity with the solid model}\label{sec:distsolid}

So far, we only described how porosity (i.e. the distention parameter $\alpha$) is 
used to modify the pressure: $$P(\rho,E) \rightarrow \frac{1}{\alpha} P_s(\rho \alpha,E).$$ In this section, 
we show how distention interfaces with the material model exposed in the previous section, which is used 
to describe the behavior of solids under strain increase.

\subsection{Distention and strength}\label{sec:diststrength}

As we have discussed in Sec.~\ref{sec:porosity}, the pressure $P$ is calculated using 
the matrix density $\rho_s$ instead of $\rho$. Consequently, the deviatoric stress tensor has to be 
modified as well. In order to compute the time evolution of $S^{\kappa\lambda}$ as a function of the matrix 
variables, we introduce the following factor:
\begin{equation}\label{eq:fdef}
 f \equiv \frac{[\vec\nabla \vec v]_s}{ [\vec\nabla \vec v]}
\end{equation}
This factor relates the velocity divergence of the matrix and of the porous material. Using the continuity 
equation for the matrix
\begin{equation}
\dot\rho_s = -\rho_s [\vec \nabla \vec v]_s
\end{equation}
and for the porous material
\begin{equation}
\dot\rho = -\rho [\vec \nabla \vec v]
\end{equation}
we can write the factor $f$ as
\begin{equation}
f=\frac{\dot\rho_s}{\rho_s}\frac{\rho}{\dot\rho}=\frac{\dot\rho_s}{\alpha \dot\rho}
\label{f}
\end{equation}
Using $\dot \rho_s=\alpha \dot \rho + \dot \alpha \rho$ we finally get
\begin{equation}
f = 1 + \frac{\dot\alpha \rho}{\alpha\dot\rho}
\label{fc}
\end{equation}
The factor $f$ is then used to compute the time evolution of $S^{\kappa\lambda}$ for the porous material:
\begin{equation}
\frac{dS^{\kappa\lambda}}{dt} \rightarrow f\frac{dS^{\kappa\lambda}}{dt}
\label{fdS}
\end{equation}
The multiplication by the factor $f$ is motivated by the fact that both, the velocity divergence 
\begin{equation}
\vec\nabla \vec v = \frac{\partial v_1}{\partial x_1} +  \frac{\partial v_2}{\partial x_2} +  \frac{\partial v_3}{\partial x_3}
\end{equation}
and the time derivative of the deviatoric stress tensor (Eq.~\ref{eq:ds}) are linear combinations of the spatial derivative of the components of the velocity vector (the linearity of Eq.~(\ref{eq:ds}) follows from Hooke's law). Since according to Eq.~(\ref{eq:fdef}), the velocity divergence of the matrix is given by 
\begin{equation}
  [\vec\nabla \vec v]_s = f [\vec\nabla \vec v],
\end{equation}
we obtain
\begin{equation}
\left[\frac{dS^{\kappa\lambda}}{dt}\right]_s = f \left[\frac{dS^{\kappa\lambda}}{dt}\right].
\end{equation}
In addition to the multiplication by $f$, the deviatoric stress tensor $S^{\kappa\lambda}$ is multiplied by  $\alpha^{-1}$ as it is done 
with the hydrostatic pressure $P$. We finally write the time evolution of $S^{\kappa\lambda}$ in the following 
form:
\begin{equation}\label{fdSa}
\frac{d}{dt}\left[\frac{1}{\alpha}S^{\kappa\lambda}\right]=\frac{1}{\alpha}\frac{dS^{\kappa\lambda}}
{dt}-\frac{1}{\alpha^2}S^{\kappa\lambda}\frac{d\alpha}{dt}
\end{equation}
where $dS^{\kappa\lambda}/dt$ is modified according to Eq.~(\ref{fdS}). 

The computation of the factor $f$ can fail for small $\dot \rho$. This is the main reason why the 
$\rho-alpha$ model was used in our first implementation (Benz and Jutzi, 2006) 
as in this case, this factor can be computed using a simpler relation: $f=(\rho/\alpha)(d\alpha/d\rho)$. For 
several reasons (see Sec.~2.4), the $P-alpha$ model is actually more appropriate. Therefore, we have worked out a functional form for the factor $f$ that does not lead to difficulties for small $\dot\rho$. For this, we replace $\dot E$ in Eq.~(\ref{dadt}) 
by $\dot E = P / \rho^2 \cdot \dot\rho$ and we rewrite Eq.~(\ref{dadt}) as
\begin{equation}
 \dot\alpha= \frac{P/\rho^2 \left(\frac{\partial P_s}{\partial E_s}\right) + \alpha  \left(\frac{\partial P_s}
{\partial \rho_s}\right)}{\alpha + \frac{d\alpha}{dP} \left[P - \rho \left(\frac{\partial P_s}{\partial 
\rho_s}\right)\right]}\cdot \frac{d\alpha}{dP} \cdot \frac{d\rho}{dt}
\end{equation}
Defining
\begin{equation}
\frac{d \alpha}{d\rho} \equiv \frac{P/\rho^2 \left(\frac{\partial P_s}{\partial E_s}\right) + \alpha  
\left(\frac{\partial P_s}{\partial \rho_s}\right)}{\alpha + \frac{d\alpha}{dP} \left[P - \rho 
\left(\frac{\partial P_s}{\partial \rho_s}\right)\right]}\cdot \frac{d\alpha}{dP}
\label{dadrhom}
\end{equation}
the derivative of $\alpha$ can be written as
\begin{equation}
 \dot\alpha =  \frac{d\alpha}{d \rho}\dot\rho
\end{equation}
and we finally compute the correction factor in the following form:
\begin{equation}
f = 1 + \frac{\dot\alpha \rho}{\alpha\dot\rho} = 1 + \frac{d\alpha}{d\rho}\frac{\rho}{\alpha}
\label{fcm}
\end{equation}
The time evolution of the deviatoric stress tensor is then computed using Eq.~(\ref{fdSa}) and (\ref{fcm}). In this way, the deviatoric stress is a function of the distention and and also of the used $P-alpha$ relation. On the other hand, we do not explicitly relate the yield strength $Y$ (used for the von Mises yielding criterion) and distention.

\subsection{Distention and damage}\label{pord}

Porosity does not only affect the stress behavior. It also has to be taken into account to compute the state 
variable damage.

\begin{figure}[h!]
\centering
\resizebox{!}{6cm}{\includegraphics{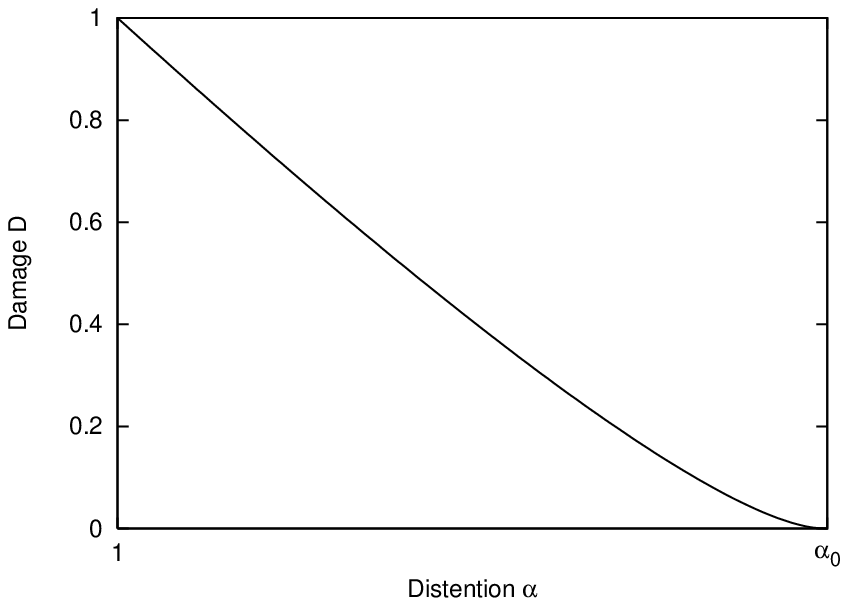} }
\caption{}
\label{fig:distdam}
\end{figure}

Compression of a porous material is accompanied, if significant enough, by the breaking of cell walls. Our model takes into 
account this crushing behavior relating distention with the state 
variable damage. Since both damage $D$ and distention $\alpha$ are defined as volume 
ratios (Eqs.~(\ref{defD}) and (\ref{defalpha}), respectively), we assume for simplicity a linear relation 
between $D$ and $\alpha$ (other forms will be investigated in the future). The conditions 
$D=0$ at $\alpha=\alpha_0$, and $D=1$ when all pores have been crushed ($\alpha=1$), lead to the following 
expression:
\begin{equation}
D = 1 - \frac{(\alpha-1)}{(\alpha_0-1)}.
\label{defad1}
\end{equation}
The time evolution of $D^{1/3}(\alpha)$ is then given by
\begin{equation}
\frac{dD^{1/3}}{dt}=\frac{dD^{1/3}}{d\alpha}\frac{d\alpha}{dt}
\label{tevD1}
\end{equation}
and using Eq.~(\ref{defad1}) we obtain
\begin{equation}
\frac{dD^{1/3}}{dt}=-\frac{1}{3}\left[1-\frac{\alpha-1}{\alpha_0-1}\right]^{-\frac{2}{3}} \frac{1}{\alpha_0-1}\frac{d\alpha}{dt}.
\label{tevDtmp}
\end{equation}
A close examination of Eq.~(\ref{tevDtmp}) reveals a problem
since for $\alpha=\alpha_0$, the derivative $D^{1/3}/dt$ becomes infinite. In order to avoid this
problem we add the small quantity $\delta D$ to the linear
relation (\ref{defad1}) and normalize it so that the conditions $D(\alpha_0)=0$ and $D(1)=1$ are still 
satisfied:
\begin{equation}
D^{1/3} \rightarrow \frac{\left(D + \delta D\right)^{1/3} -(\delta D)^{1/3}}  {\left(1+\delta D\right)^{1/3}-
(\delta D)^{1/3}}
\end{equation}
Using Eqs.~(\ref{tevD1}) and (\ref{defad1}) we finally get
\begin{equation}\label{tevD2}
\frac{dD^{1/3}}{dt}=-\frac{1}{3}\frac{\left[1-\frac{(\alpha-1)}{(\alpha_0-1)} + \delta D 
\right]^{-\frac{2}{3}}}{\left(1+\delta D\right)^{1/3}-(\delta D)^{1/3}}\frac{1}{(\alpha_0-1)}
\frac{d\alpha}{dt}
\end{equation}
where we set $\delta D=0.01$. The actual relation between distention and damage is shown in Figure \ref{fig:distdam}.

We now have two equations describing damage growth: the first treats damage under tension (\ref{Dgr}) while the 
second (\ref{tevDtmp}) is related to the compression of the (porous) material. Note that we do not \textit{explicitly} model damage increase due to shear deformation. However, since the local scalar strain is computed from the maximum negative stress after principal axis transformation (Eq.~\ref{eq:epsi}), shear fracturing is implicitly accounted for in our model.

In order to get the total growth of damage, we build the sum of the two differential equations:
\begin{equation}
\left[\frac{dD^{1/3}}{dt}\right]_{total}=\left[\frac{dD^{1/3}}{dt}\right]_{tension}+\left[\frac{dD^{1/3}}{dt}
\right]_{compression}
\end{equation}
We now use this equation instead of Eq.~(\ref{Dgr}) to compute damage $D$ which varies between 0 and 1. According to Eq.~(\ref{eq:sd}), damage leads to a reduction of both tensile and shear strength by a factor of $(1-D)$.

According to our model, damage can grow only. We do not include any restoration of damage as Sirono (2004) proposed for low density grain aggregates.

\subsection{Material parameters}

All  parameters used by our porosity model are material parameters which can in principle be measured  
experimentally. Some of these parameters, such as the crush-curve, can be measured more easily than others (such as Weibull parameters, shear and tensile strengths). Unfortunately, such measurements are rarely done in 
practice and we plan to carry out some of them for several porous materials in future works. 

The lack of an experimentally determined reliable database of relevant material parameters is actually one of 
the most limiting factor in our model. In particular, the thorough testing of the model by comparison with 
experiments is rendered particularly difficult if all material properties have not been measured properly. 
Freely choosing the missing values so as to match an experiment is not a satisfactory approach for an ab initio 
method such as ours. Unfortunately, this is often the only alternative we have.
However, recent experiments have been performed on some porous materials, and material properties 
have been measured. A detailed comparison between results of simulations using our model with these 
experiments will be the subject of a next publication.

\section{Tests}\label{sec:tests}

In this section we present some simple test cases where we compare our numerical model to 
expected theoretical solutions. We also present a Deep Impact-like simulation to show the effect of porosity on the outcome of an impact. Further, different compaction models are compared by performing an impact in (highly) porous basalt.

The present tests aim at a first actual verification of the model by showing that 
it is consistent and correctly implemented in our code. A more detailed validation by comparison with actual 
impact experiments on porous material will be presented in a forthcoming paper. 

\subsection{1D compaction wave}\label{sec:1dwave}

As a first numerical test we consider a (plane) shock wave travelling in only one spatial
dimension. We compare the simulation with the analytical results obtained by solving the corresponding Hugoniot 
equations.

In this simulation, we use porous aluminium with an initial distention of $\alpha_0=1.275$. We use a 
$P-alpha$ relation with both an elastic and a plastic regime. In the elastic regime $P<P_e$, the distention is computed using Eq.~(\ref{dadpe}). Using this equation instead of $d\alpha/dP=0$ leads to a smaller velocity of the elastic wave. We actually use Eq.~(\ref{dadpe}) in this case because it provides an additional test (properties of the elastic wave). However, for all other simulations presented in this paper we assume that $\alpha_e=\alpha_0$ and therefore $c_e=c_0$.

For the plastic regime we use the quadratic relation (\ref{alphaqudratic}). The porosity parameters 
used in this simulation are given in Table \ref{porosityalu}.

As for the equation of state, we use the (full) Tillotson equation with aluminum parameters (Melosh, 1989). Since we are 
dealing with a purely one dimensional problem, strength is not included in this simulation and only the 
diagonal part of the stress tensor (pressure) is considered. 

The analytical solution of this problem can be obtained by solving the Hugoniot equations together with 
Eq.~(\ref{meos}) and the $P-alpha$ relation defined above. 

To carry out the simulation we use a cylinder aligned along the z-axis with a
radius $r=0.2$ cm and a height $h=2$ cm. Since we want to study a one dimensional case, the forces acting 
on the particles in the x-y plane are set to zero:
$$
\left(\frac{dv_x}{dt}\right)_i=\left(\frac{dv_y}{dt}\right)_i = 0
$$
The shock wave is then produced by moving the particles of the first layer (at the top of the cylinder) with a 
constant velocity $v_z$ = -45.8 $\times$ 10$^{3}$ cm/s in the z-direction. The last layer (at the bottom of the cylinder) is 
fixed. We use $5.6 \times 10^5$ particles in this simulation. However, in order to reduce boundary effects, only
particles in a cylinder of $0.05$ cm radius are used for the comparison with the theoretical solution.

\begin{figure}[h!]
\centering
\resizebox{10cm}{!}{\includegraphics{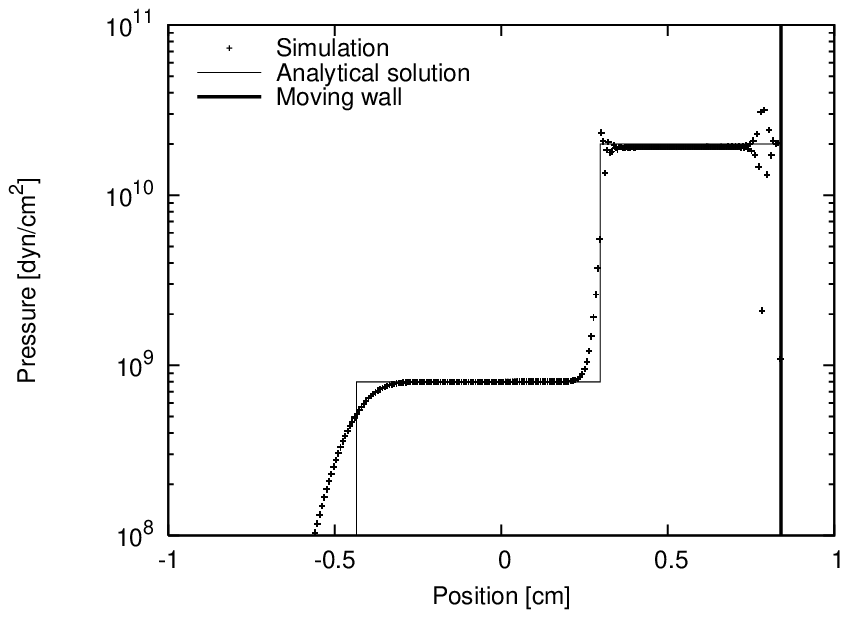} }
\vspace{1cm}
\resizebox{10cm}{!}{\includegraphics{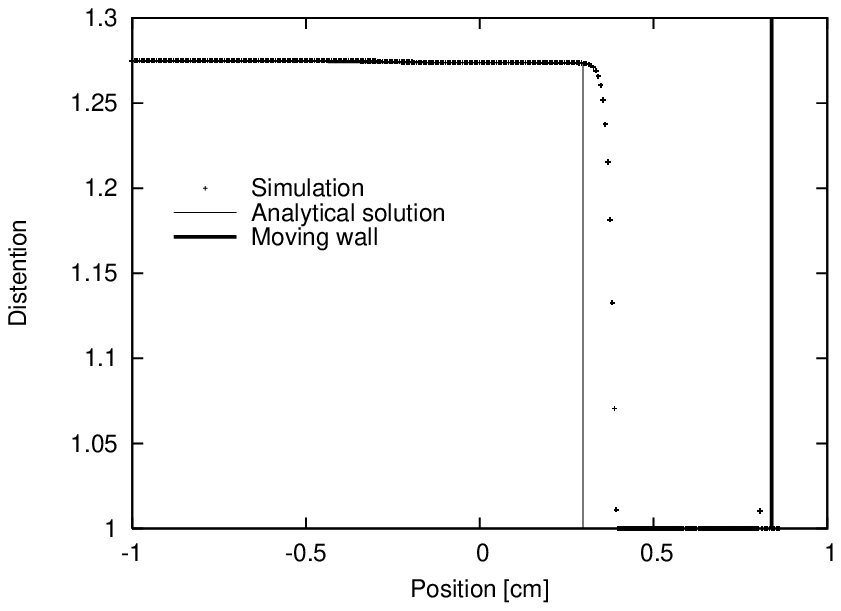} }
\caption{}
\label{fig:zpza}
\end{figure}

Figure \ref{fig:zpza} (top) shows the compaction wave travelling in the
z-direction at time $t$ = 3.5 $\mu$s. As expected, there are two waves: first, 
there is a so-called elastic precursor with an amplitude equal to the elastic
pressure $P_e$. The elastic precursor is followed by a plastic
compaction (shock) wave. The distention is only slightly changed due to the elastic wave and is decreased to $\alpha=1$ by the compaction wave (see Fig.~
\ref{fig:zpza}, bottom).

There is a good agreement between the theoretical solution (solid line) and the 
velocities of the two waves (the elastic precursor and the shock wave) obtained by the simulation.
Moreover the corresponding pressure amplitudes have the expected value. The oscillations behind the compaction wave are due to the moving layer of particles (wall). The smoothed shape of the wave (especially of the elastic precursor) is caused by the artificial viscosity and the smoothing due to 
the SPH interpolation. No effort was made to fine tune the artificial viscosity to achieve a less diffusive solution.

\subsection{Crush-curve simulation}

As we have discussed in Sec.~\ref{sec:porosity}, an advantage of defining the 
distention as a function of pressure is that this relation can be directly determined experimentally by 
compressing a porous sample. Such an experiment (Nakamura et al., in preparation) was performed by A.M. Nakamura and K. Hiraoka at Kobe University (Japan). In this experiment, a cylindrical sample of porous pumice, confined within a steel cylinder, is 
compressed in one dimension. The applied force $F$ and the corresponding displacement $e$ of the penetrating 
piston (which defines the actual length $l$ of the sample) are measured. From these quantities, the applied 
pressure $P_m$ is given by the force $F$ per unit area and the actual distention can be determined in the 
following way
\begin{equation}\label{eq:alpham}
 \alpha_m = \alpha_0 \frac{l}{l_0}
\end{equation}
were $l_0$ is the initial length of the sample and $l=l_0-e$ is the actual length. In this way, we can 
obtain the relation $\alpha_m(P_m)$. Using the function (\ref{alpha2reg}), we fit the curve $\alpha_m(P_m)$ and get the required analytical relation $\alpha=\alpha(P)$. The fitting parameters are given in Table \ref{porositypumice}.

As a further test, we use this relation in our porosity model and we simulate the compaction experiment.
The simulation is performed moving the first layer of particles of a cylinder $A$ aligned along the $z$-axis with 
the velocity $v_z$.
As in the experiment, the walls and the bottom of the cylinder are fixed. In order to determine the actual 
(macroscopic) distention and the applied pressure, we
define a test cylinder $B$ within $A$ with a radius $r$
and an initial length of $l_0$ corresponding to the sample in the experiment. Since it is a quasi--static 
compression, we assume that the applied pressure corresponds to the pressure in the sample and we compute 
$P$ as the average pressure in the cylinder $B$. As in the experiment, we define the actual distention using equation Eq.~(\ref{eq:alpham}). 

\begin{figure}[h!]
\centering
\centerline{\epsfig{file=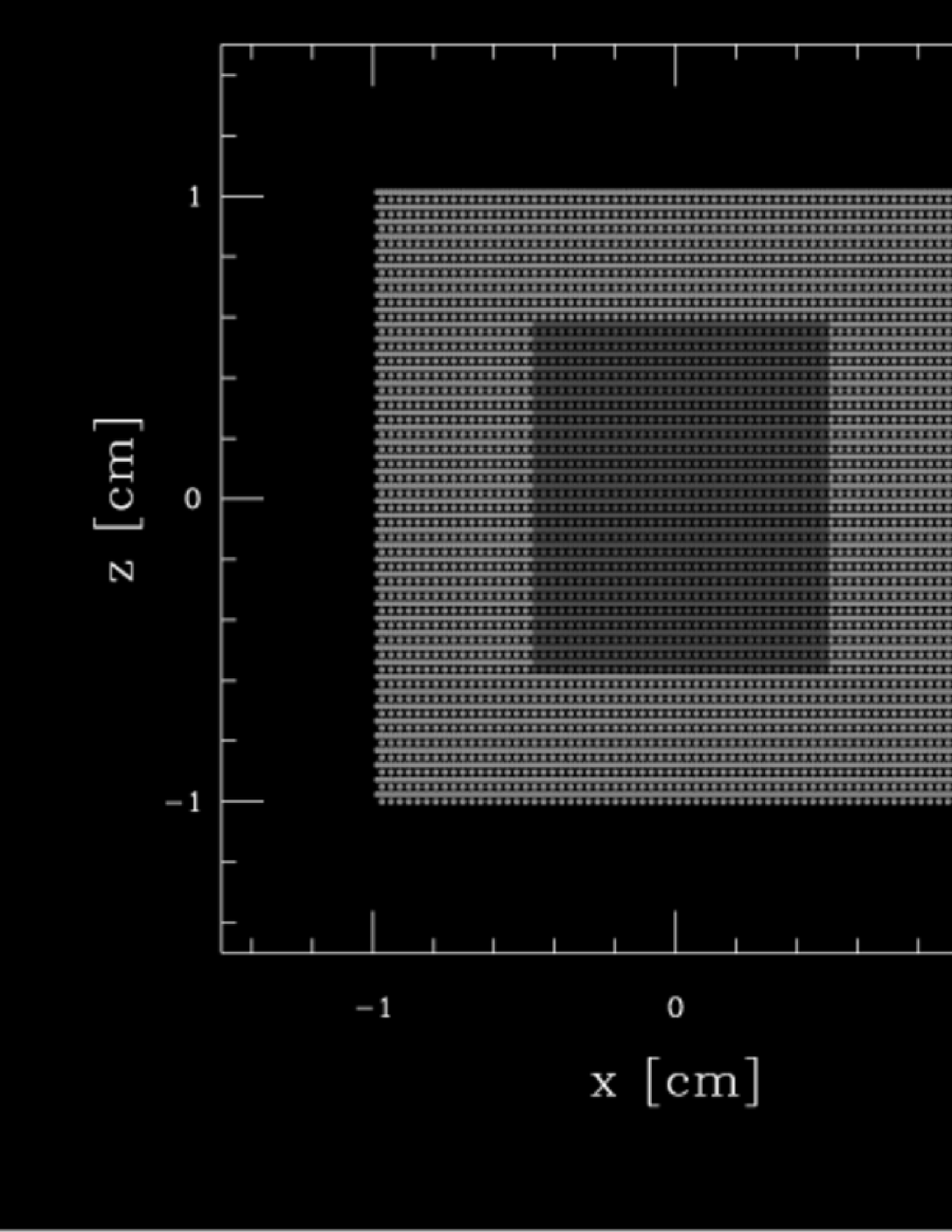,height=8cm}}
\vspace{1. truecm}
\centerline{\epsfig{file=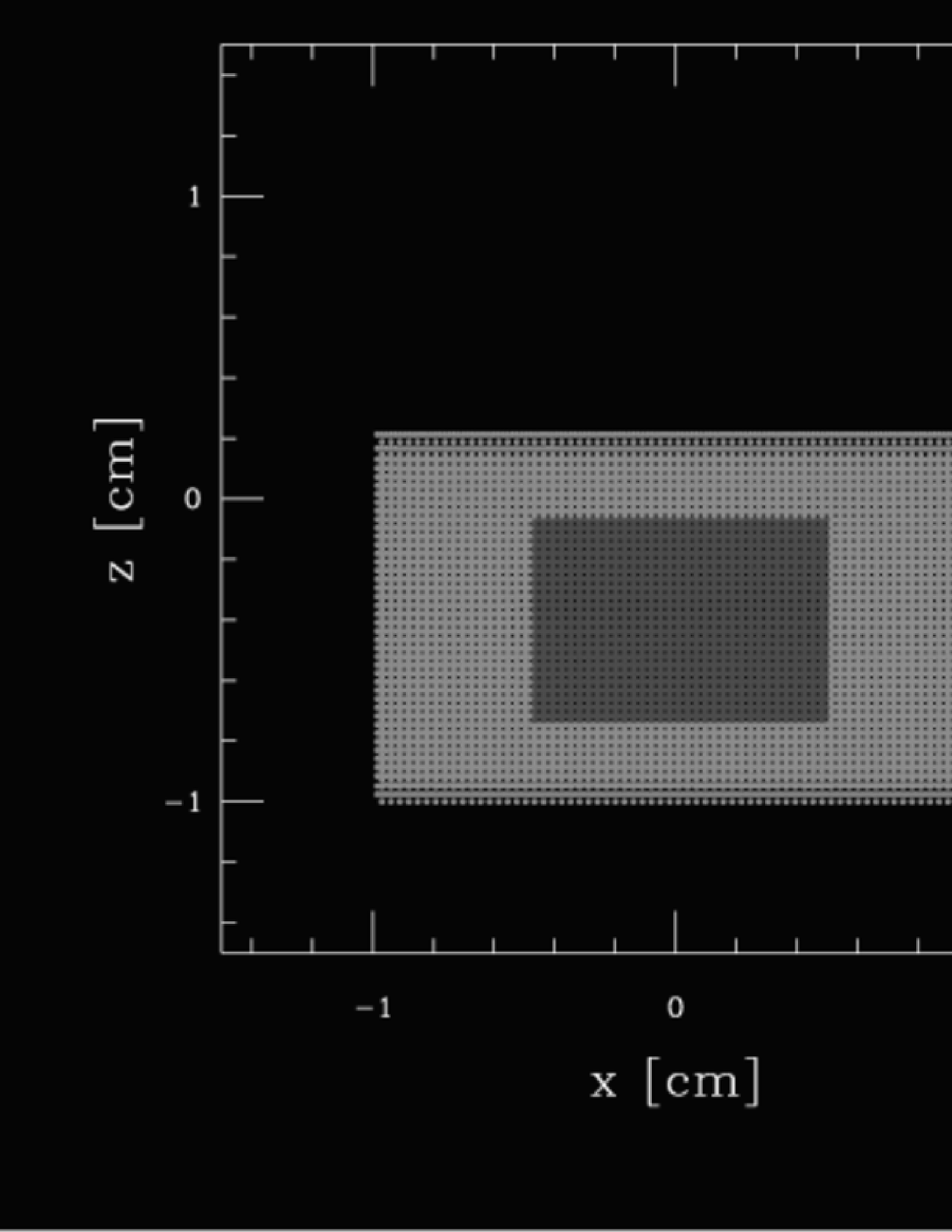,height=8cm}}
\caption{}
\label{fig:ccsim}
\end{figure}

Figure \ref{fig:ccsim} illustrates the setup of the simulation which has been performed in 3D. On this 2D 
slice, the dark particles correspond to the cylinder $B$ which represents the sample. The simulation is shown at 
the initial state (top) and during the compression (bottom). In order to control the uniformity of the compaction, we compute the standard deviation of the distention in the sample (particles within $B$) at certain timesteps. During the whole simulation, the standard deviation is of the order of 1\% which indicates that the compaction proceeds uniformly.

\begin{figure}[h!]
\centering
\resizebox{12cm}{!}{\includegraphics{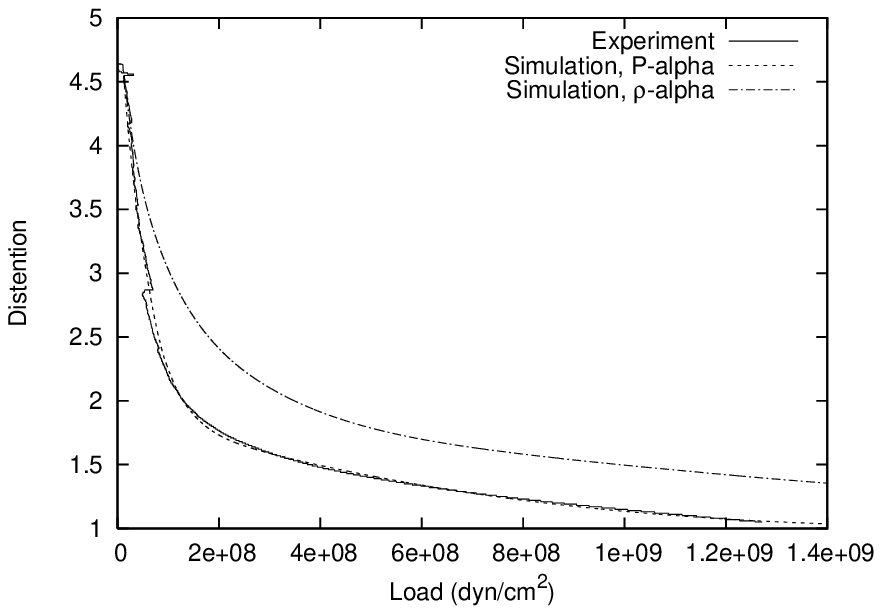} }
\caption{}
\label{fig:cccomp}
\end{figure}

In Figure \ref{fig:cccomp}, the crush-curve obtained by the simulation is compared to the experimental 
crush-curve. There is a very good agreement between the two curves. Of course, using the measured crush-curve as an input, one expects to obtain the same curve by the simulation. However, we have to point out that the pressure - distention relation used as input in the code is not necessarily the same as the pressure - distention relation measured in the simulation. In the $\alpha(P)$ relation used in the code, distention is defined as $\alpha=\rho_s/\rho$. In the relation obtained by the simulation, distention is defined according to Eq.~(\ref{eq:alpham}), which can be written as: 
\begin{equation}
 \alpha_m = \alpha_0 \frac{l}{l_0} = \frac{\rho}{\rho_{s0}}
\end{equation}
This definition therefore assumes a constant matrix density ($\rho_s=\rho_{s0}$). As we discussed in Sec.~\ref{sec:highpor}, the matrix density changes during the compaction. However, the change of $\rho_s$ is only small (typically 1\%) leading to an error of $\alpha_m$ which is in the same order. The fact that the two curves (simulated crush-curve and the crush-curve used as input) are in a very good agreement indicates that the change of the matrix denstiy is indeed very small. This result therefore suggests that the distention defined by $\alpha_m=\rho_{s0}/\rho$ is a reasonable approximation of the true distention $\alpha=\rho_s/\rho$ (even for highly porous material). However, it is important to note that even though the matrix density changes only very little, it can still have a great effect (see Sec.~\ref{sec:highpor}).

In a quasi static compression, the energy increase ($P dV$ work) is generally lower than in the case of shock compression, leading to a smaller thermal pressure.  In the simulation above we assumed that the system is thermally insulated (i.e., the energy radiation is not taken into account). Therefore, the thermal pressure can still not be ignored. In Fig.~\ref{fig:cccomp}, the crush-curve obtained by a simulation using a relation $\alpha=\alpha(\rho)$ which follows from the $\alpha = \alpha(P)$ curve neglecting the thermal pressure ($\rho - alpha$ model) is shown. To reach a given distention, a higher pressure is needed using the $\rho - alpha$ model instead of the $P - alpha$ one. This difference corresponds to the thermal pressure.

\subsection{Impact simulations}
In this section, we show several impact simulations using different material types (ice and basalt) and compaction models (no porosity model; $P - alpha$, $\rho - alpha$ and $\epsilon - alpha$ model). Note that the following results were obtained using a pre-damaged (i.e. strengthless) material. Therefore, we are in the gravity regime where the final crater size and the total amount of ejected material depend on the gravity acceleration. Due to the high resolution and the resulting small timestep (which is much smaller than the crater formation time), we only simulate the first phase of the crater formation.
\subsubsection{Deep-impact like impact} \label{sec:deepimp}
As a first application of our model, we show the effect of porosity on the outcome of an impact event by comparing simulations using porous and non-porous targets. Since we want to model a realistic case, we simulate a Deep-Impact-like impact, inspired from the Deep Impact Space mission (A'Hearn and Combi, 2007). 

We model the target in two different ways. In the first case, we use an initial distention of $\alpha_0=3.0$ and we explicitly model porosity. To simplify matters we use a quadratic relation (Eq.~\ref{alphaqudratic}) for $\alpha=\alpha(P)$ with $P_e =$ 1$\times$10$^7$ dyn/cm$^2$ and $P_s =$ 2$\times$10$^9$ dyn/cm$^2$. These values are chosen rather arbitrarily, however, the resulting $\alpha(P)$ relation looks reasonable compared to measured crush-curves of other materials (see for expample Fig.~\ref{fig:cccomp}).
In the second case, we use the same initial density ($\rho_0=\rho_{s0}/\alpha_0$), but the target is modeled as a solid (without porosity model).
As target material we use pre-damaged (strengthless) ice. Only a small part of the target (comet Tempel-1) is modeled (half sphere with a radius of 28 m). In order to have a reasonable resolution in the region of interest (crater), we use a rather high number of particles for the target ($N_t$ = 5.2$\times$10$^6$). The resulting mass per SPH particle is $m_p = 2.6$ kg and the smoothing length is $h = 23$ cm.
 The impactor is modeled as a 370 kg aluminium sphere impacting at an angle of 30 degrees (from horizontal) with a velocity of 10 km/s. We use 140 SPH particles for the impactor in order to have the same mass per particle as in the target. The Tillotson equation of state (without simplifications) is used for these simulations. 

\begin{figure}[h!]
\centerline{\epsfig{file=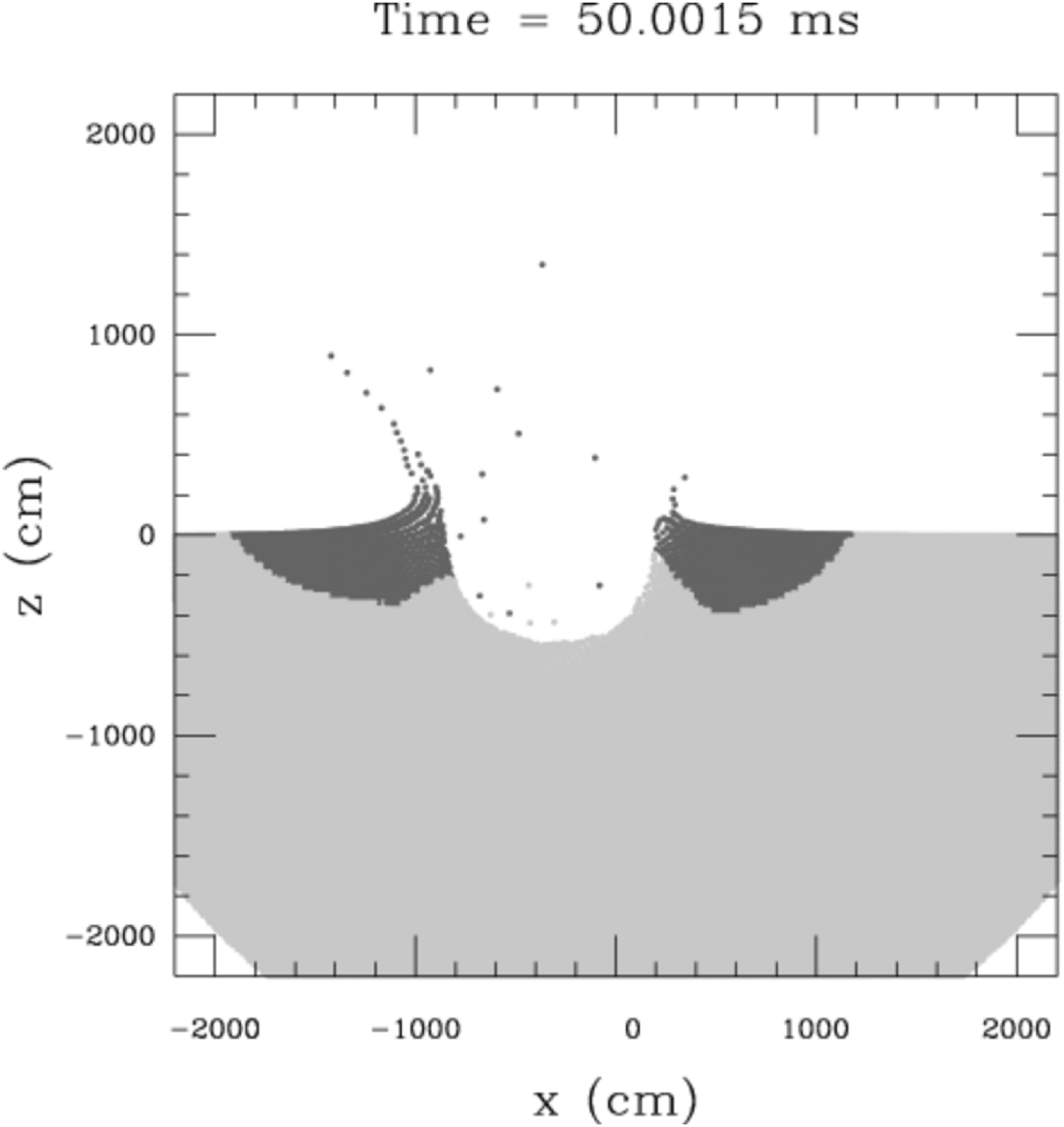,height=8cm}}
\vspace{1. truecm}
\centerline{\epsfig{file=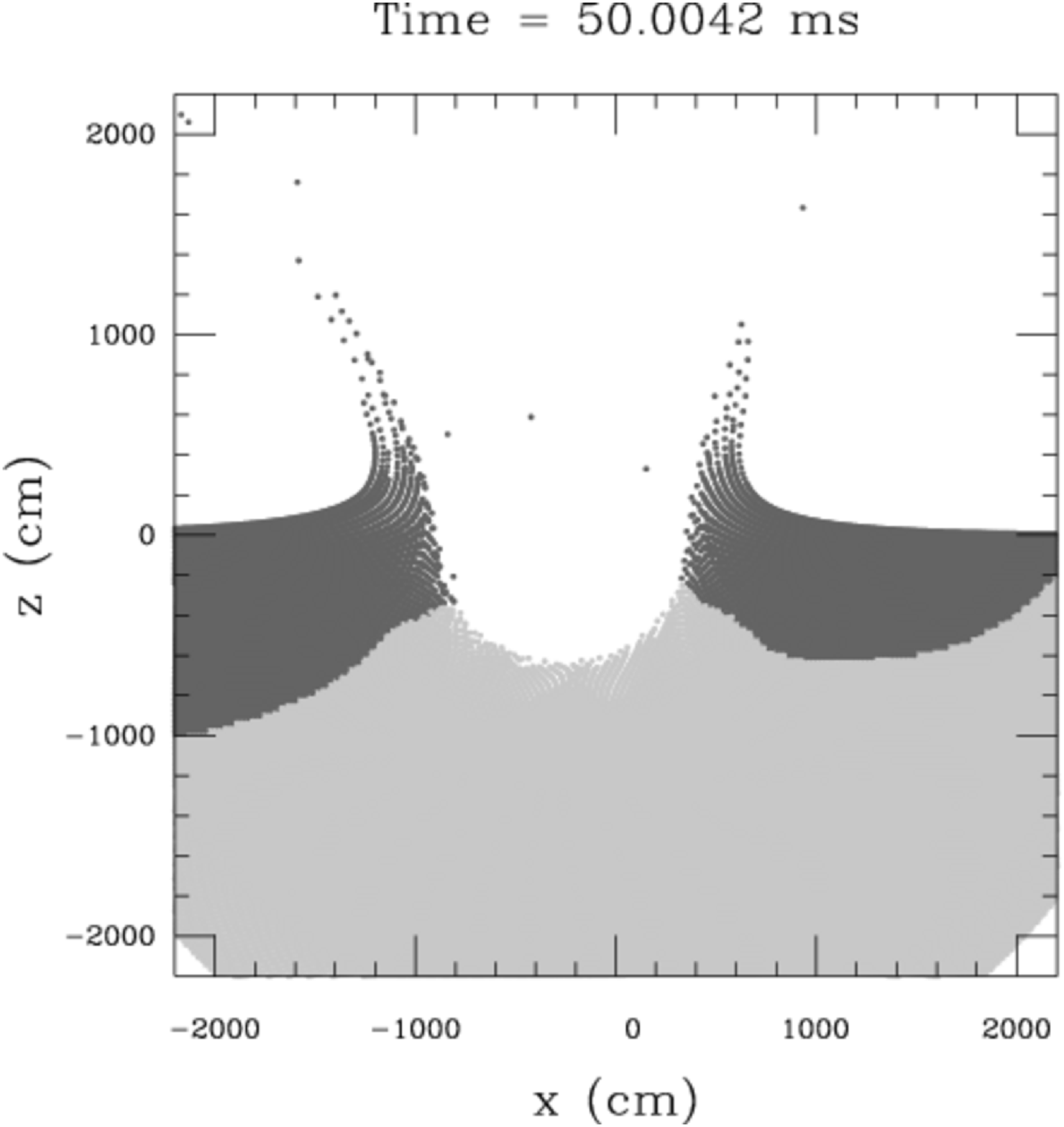,height=8cm}}
\caption{}
\label{fig:diplot}
\end{figure}

\begin{figure}[h!]
\centering
\resizebox{12cm}{!}{\includegraphics{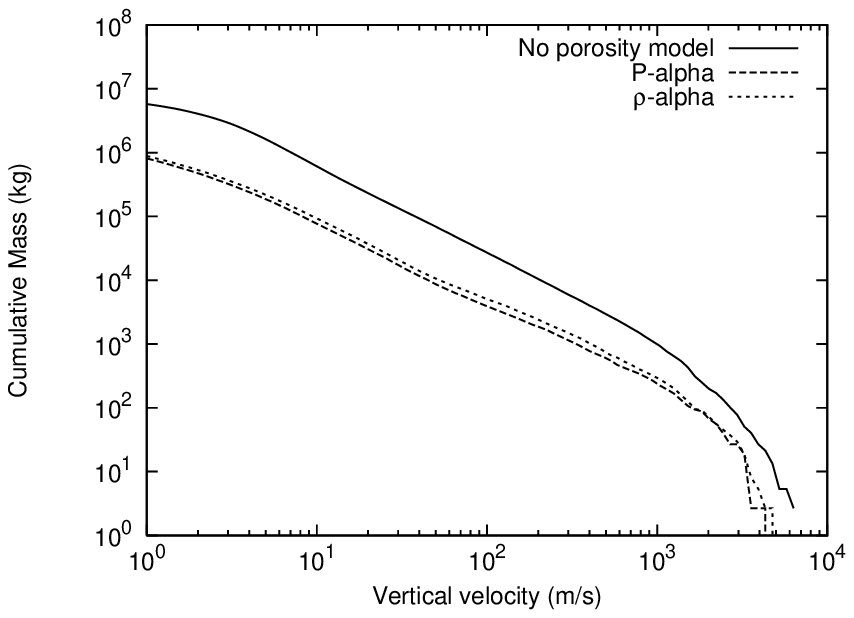} }
\caption{}
\label{fig:divel}
\end{figure}

Figure \ref{fig:diplot} shows the outcome of the simulation after 50 ms in two dimensional slices of the three dimensional target. The dark particles in these plots have vertical velocities greater than 2 m/s (which is about the escape velocity of Tempel-1). Obviously, there is much less material ejected in the case where we explicitly model porosity (top) than in the solid target simulation (bottom). In fact, the difference of ejected mass is about a factor ten. Figure \ref{fig:divel} shows the amount of mass ejected with a velocity higher than a certain velocity in both cases.

For comparison, the results obtained by using the $\rho-alpha$ model (instead of $P-alpha$) are also plotted in Fig.~\ref{fig:divel}. For this, we use a relation $\alpha=\alpha(\rho)$ which follows from $\alpha=\alpha(P)$ by neglecting the thermal component of the pressure (see Sec.~\ref{sec:rapa}). There is only a very small difference between the results of the two models. This is consistent with the estimation (Eq.~\ref{eq:acrit}) of $\alpha_{0crit} \simeq 6$ for $c$ = 0.4 (ice). Since $\alpha_0 = 3 < \alpha_{0crit}$, full compaction is  possible and since $c$ is very small, we do not expect big differences. However, the situation looks different if we use basalt instead of ice as target material.

\subsubsection{Comparison between different compaction models}

For the typically used basalt parameters we get $c \simeq 2$ and therefore $\alpha_{0crit} \simeq 2$. Consequentely, full compaction (at moderate pressures) is not possible for $\alpha_0 = 3 > \alpha_{0crit}$ using a compaction model which does not allow an expansion of the matrix. 
 
\begin{figure}
\centerline{\epsfig{file=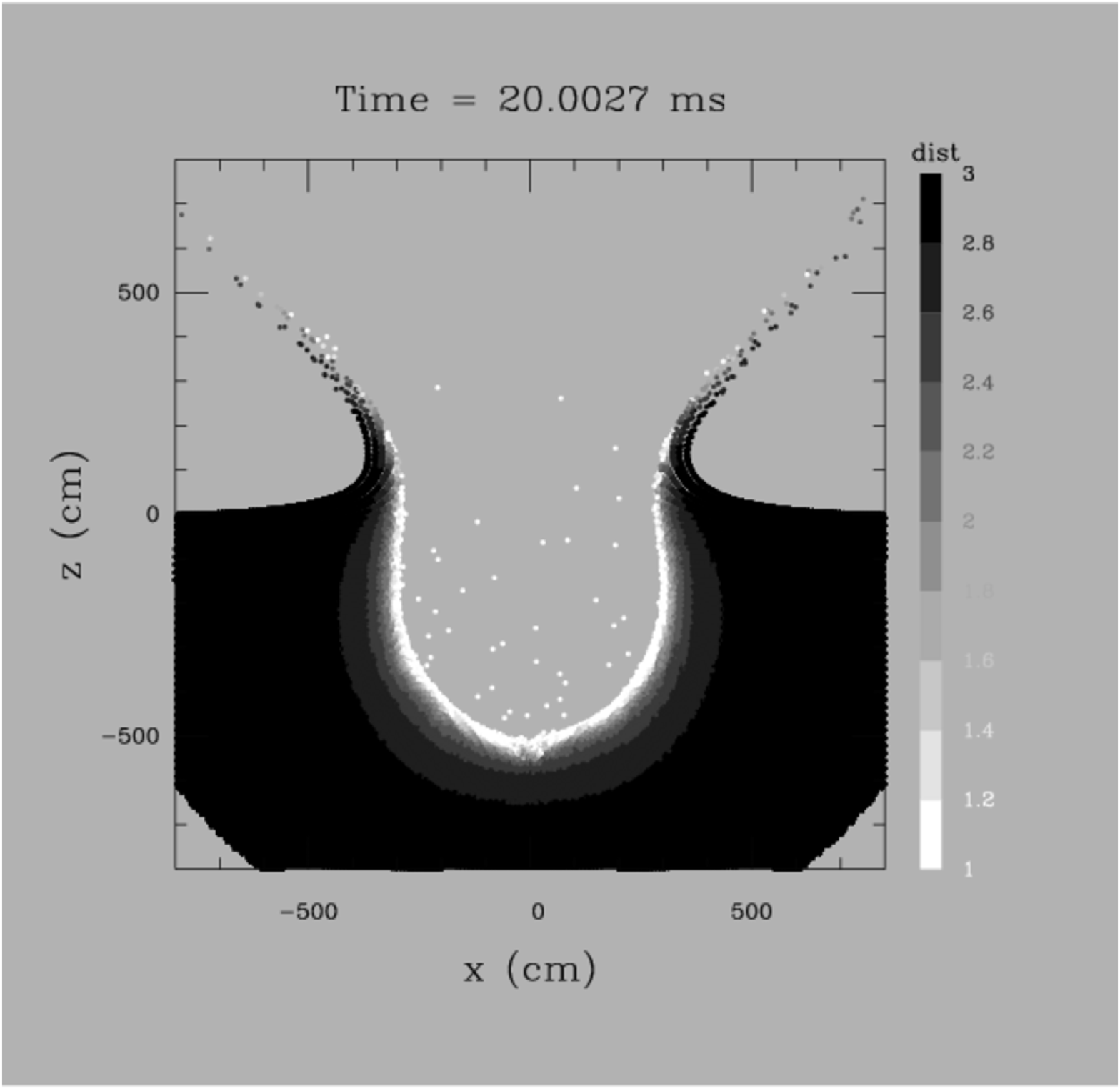,height=7cm}}
\vspace{.5 truecm}
\centerline{\epsfig{file=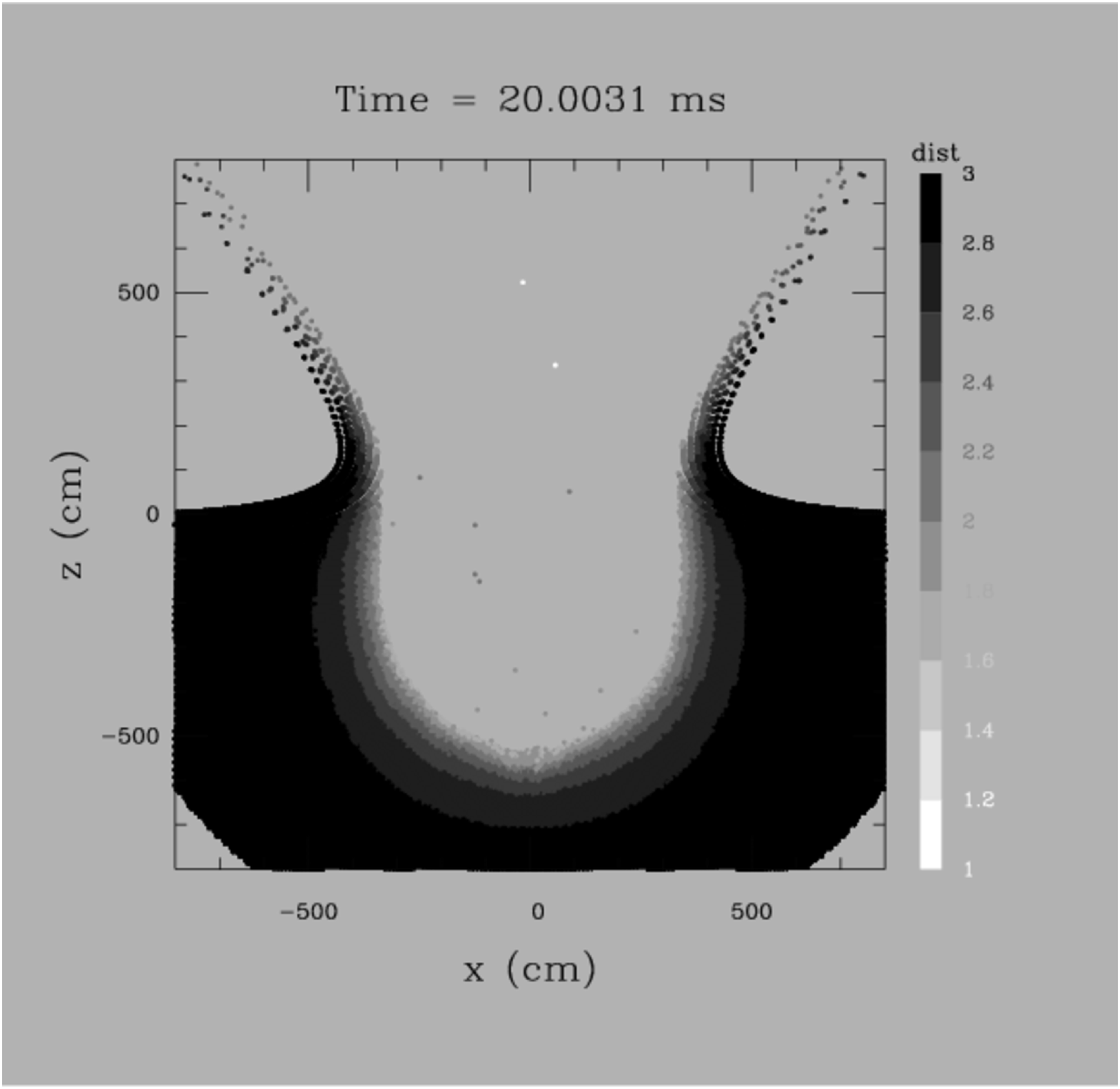,height=7cm}}
\vspace{.5 truecm}
\centerline{\epsfig{file=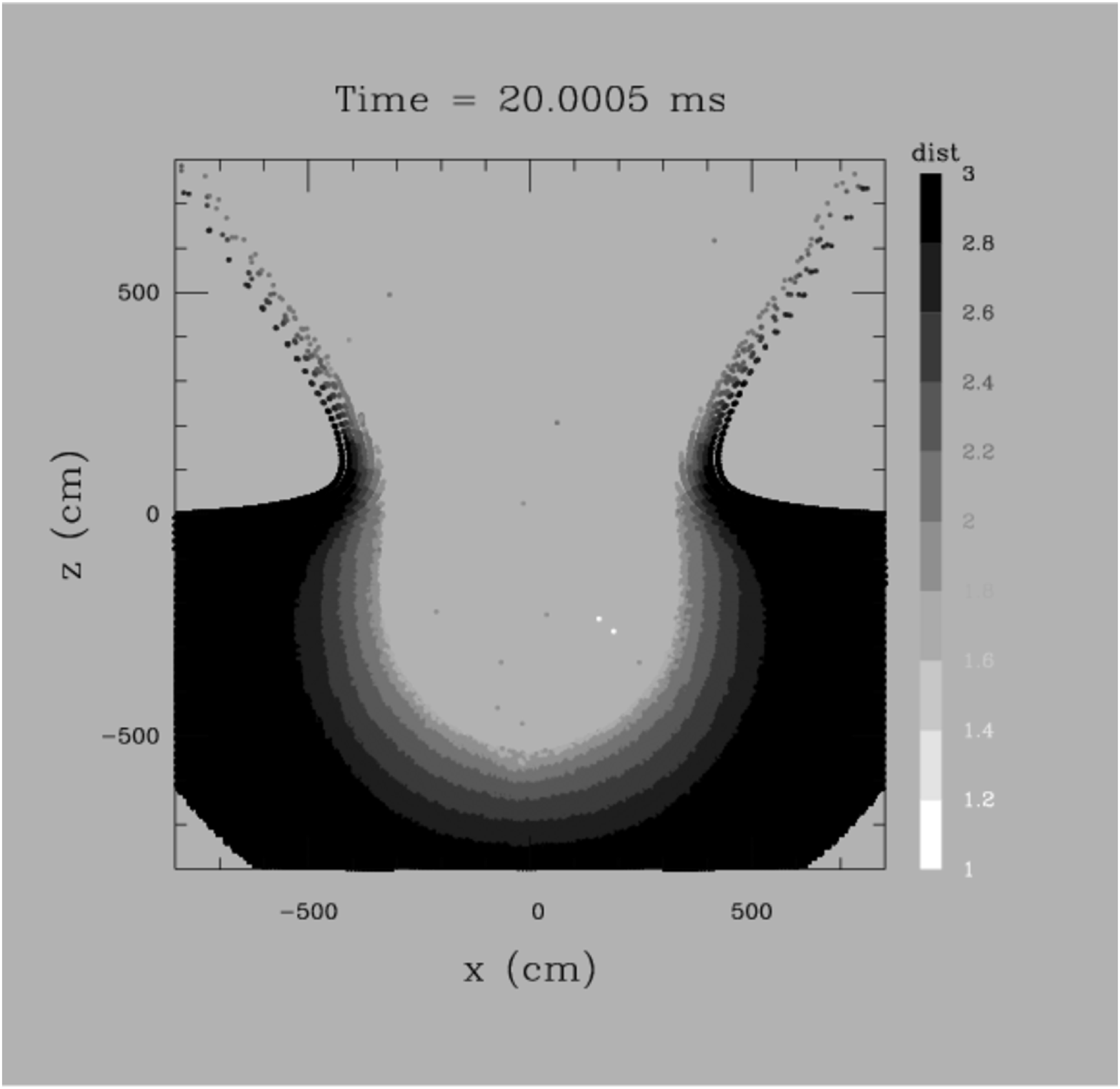,height=7cm}}
\caption{}
\label{fig:dibasalt}
\end{figure}

To illustrate the different compaction behaviors using different compaction models, we simulate an impact in a basalt target with  $\alpha_0=3.0$.  For this simulation, we use the same initial conditions as above but we use a smaller target and the impact is now head on.
We investigate three cases using the follwing compaction models:
\begin{enumerate}
 \item $P - alpha$ with the same parameters as above.
 \item $\rho - alpha$, using the relation $\alpha=\alpha(\rho)$ which follows from $\alpha=\alpha(P)$ neglecting the thermal pressure
 \item $\epsilon - alpha$, $\kappa = 1$ (constant matrix density assumption)
\end{enumerate}
 In Fig.~\ref{fig:dibasalt}, the degree of compaction (i.e., the distention $\alpha$) of the target material due to the impact is shown for  simulations using the model 1 (top), model 2 (middle) and model 3 (bottom). On these plots, the black color corresponds to a distention of $\alpha=\alpha_0$ and the white color to full compaction $\alpha=1$. The results are shown at a time (20 ms) where compaction is finished and the distention $\alpha$ does not change anymore.

\begin{figure}[h!]
\centering
\resizebox{12cm}{!}{\includegraphics{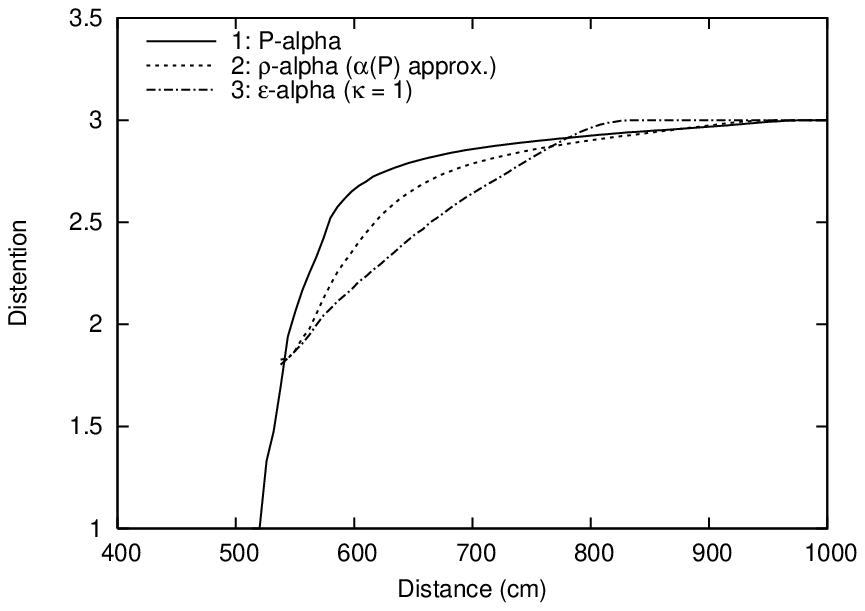} }
\caption{}
\label{fig:diprofc}
\end{figure}

As it can be seen, only in the simulation using model 1 there is a small fully compacted zone below the crater. Fig.~\ref{fig:diprofc} shows the corresponding distention profiles. The minimal distention reached by model 2 and 3 is in both cases about $\alpha$ = 1.8. This indicates that also the maximal density is about the same in both cases, because $\alpha = \rho_s/\rho$ and $\rho_s$ changes only very little (model 2) or is even constant (model 3). 
The total volume which is compacted (derived from the actual distention) is 62.3/61.9/82.9 m$^3$ for the models 1/2/3. This means that even though model 3 does not lead to full compaction, the total volume of compacted material is higher than using model 1. The reason is that the distention profile resulting from model 3 is much less steep than using model 1 (Fig.~\ref{fig:diprofc}). 

\begin{figure}[h!]
\centering
\resizebox{12cm}{!}{\includegraphics{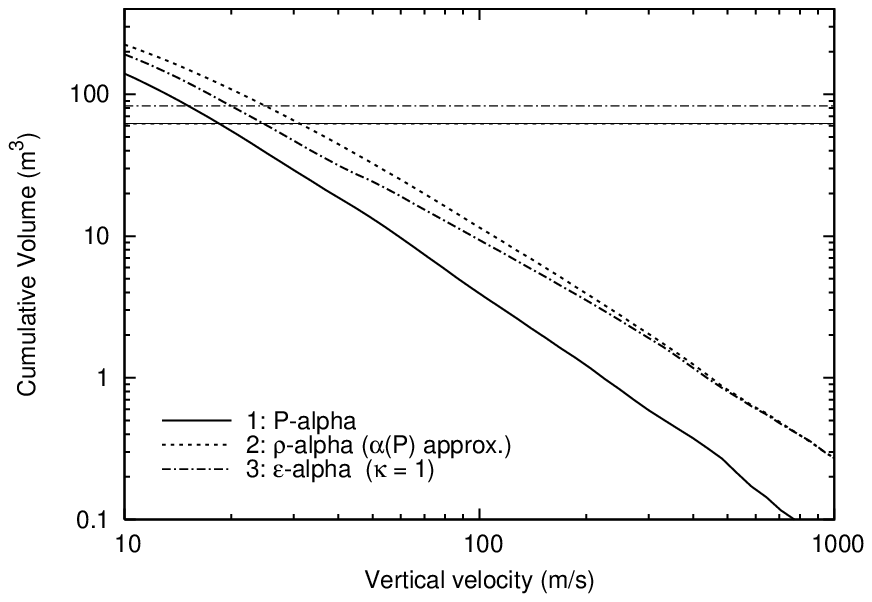} }
\caption{}
\label{fig:divvzc}
\end{figure}

In Fig.~\ref{fig:divvzc} the cumulative volume of ejected material as a function of the vertical velocity is shown. The horizontal lines in the plot represent the total compacted volume. Interestingly, using model 1, there is less ejection than using model 3 (even though model 3 leads to more compaction). This can be explained by the fact that a given pressure amplitude leads to a higher degree of compaction using model 1 than using model 3, and therefore, more energy is dissipated. Using model 2, we get the same amount of ejected volume at high velocities as with model 3, but there is more material ejected at low velocities.

\begin{figure}[h!]
\centering
\resizebox{12cm}{!}{\includegraphics{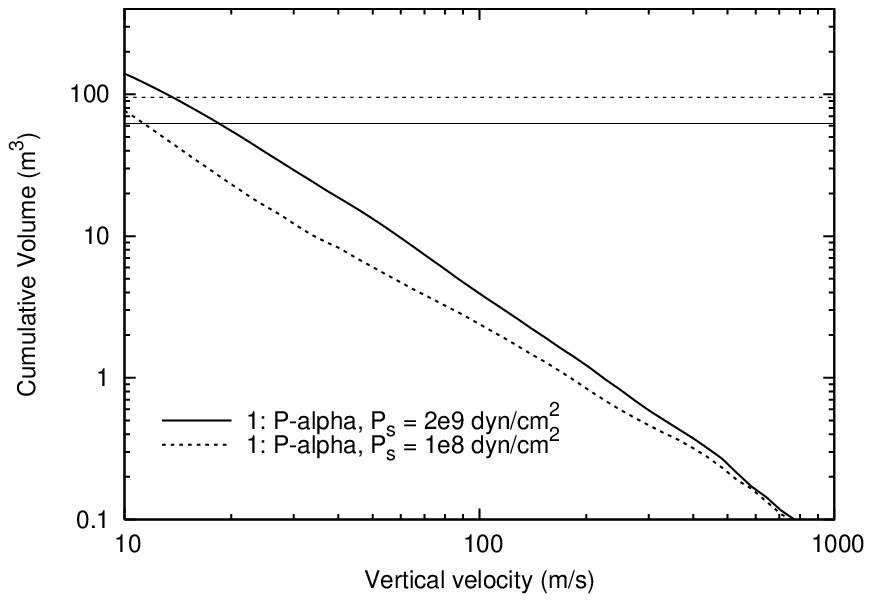} }
\caption{}
\label{fig:divvzcps}
\end{figure}

In order to study the dependence of the compacted and ejected volume on the compaction parameters, we perform a simulation using model 1 with $P_s =$ 1$\times$10$^8$ dyn/cm$^2$ instead of $P_s =$ 2$\times$10$^9$ dyn/cm$^2$. The other parameters are the same as above. Fig.~\ref{fig:divvzcps} shows the compacted volume and the cumulative volume of ejected material for the two simulations using model 1 (each with a different $P_s$).
Clearly, using a very low value for $P_s$ (1$\times$10$^8$ dyn/cm$^2$) leads to much more compaction (95.2 m$^3$) and less ejection than using $P_s =$ 2$\times$10$^9$ dyn/cm$^2$. There is even more compaction than from model 3 which assumes a constant matrix density.

\section{Conclusions}\label{sec:concl}

In this paper, we have presented a new approach to model sub-resolution porosity in
brittle solids that can be coupled to a 3D SPH hydrocode in order to simulate impacts
and collisions involving porous bodies. Such bodies are believed to be present in all the 
populations of small bodies in our Solar System. Therefore, understanding their impact response is crucial 
to determine the collisional evolution of those populations during all stages, and to assess the accretion 
efficiency of small bodies during planetary formation. Moreover, this can help defining efficient 
mitigation strategies against the impact of a porous body on Earth.

In practice, the implementation of our model does not consume excessive CPU time and is easily implemented 
in a parallel code (porosity is a local property) so that simulations involving multi-million particles can 
readily be performed. In fact, extensive testing has shown that high resolution is really needed to
obtain converged solutions in the case of simulations involving fracturing and/or porosity.

We presented two test cases. In the first test, we compared the simulation of a 1D compaction wave in porous 
material with the theoretical solution given by the Hugoniot equations. The amplitude and velocity of 
the resulting waves (elastic precursor and shock) in the simulation agree with the 
theoretical solution. The second test is a simulation of an experiment consisting in compressing a porous sample. 
We show that an advantage of the $P-alpha$ model is that an experimentally measured crush-curve can be 
directly used to define the relation $\alpha(P)$. Therefore, if a crush-curve has been measured 
for a given material, we can then simulate an impact on this material using its very crush-curve and not an 
arbitrary one which may lead to different outcomes.

As an application of our model, we presented the simulation of a Deep Impact-like impact. 
The main conclusion of this first application is that to model the behavior of a porous material 
during an impact, it is not enough to use the small bulk density expected for a porous material and 
compute its response from an impact by using a classical model of brittle failure of non-porous rock material, 
as one might be tempted to do. Actually, the cratering or disruption of porous material involves different 
processes than the ones involved in a non-porous brittle material (e.g. the crushing of pores), and this makes 
a huge difference in the outcome.

We also investigated the compaction behavior in an impact in basalt using different compaction models. We showed that using the $\rho-alpha$ or $\epsilon-alpha$ models (without modifications) can lead to difficulties to reach full compaction ($\alpha=1$) in the case of highly porous material. Using these models, the compaction behavior depends much on the EOS parameters which relate energy and pressure. On the other hand, the $P-alpha$ model can be used without modifications to simulate the compaction of highly porous material and the compaction behavior is not sensitive to these EOS parameters.

Our next step will be to validate our code in more details and in a context adapted to its future applications 
by confronting it with real high-velocity impact experiments on 
porous targets. Such a validation will mainly consist of reproducing the size distribution of the fragments 
(and their velocities when they are measured), using as inputs all measured material parameters. Indeed, 
it is important that matching the results does not rely on a fine tuning 
of all parameters but rather on the validity of the physical model. This will be the subject of a next paper.
Comparisons with impact experiments in laboratory will have to be performed in the long run for a wide 
variety of materials (porous and non porous) to improve our understanding of the impact process as a function of 
material properties and impact conditions. The formation and evolution of planetary systems involves bodies with 
a wide range of properties and colliding with each other in different regimes of impact energies, leading 
either to their accretion and the formation of planets, or to their disruption as in the current stage of our 
Solar System. 
It is then important to have the possibility to simulate the collisional process in these different regimes and 
between bodies with different degrees of porosity. Collisional evolution models, which need constraints to 
characterize the outcomes of collisions, will certainly benefit from these investigations.

\newpage
\section*{Appendix}
\subsection*{Simplified Tillotson EOS}

First, we assume that the energy involved does not exceed the energy of incipient vaporization ($E<E_{iv}$). In this state, the EOS has the following form:
\begin{equation}\label{eostpc}
     P = \left(a+\frac{b}{E/\left[E_0\eta^2\right]+1}\right)\rho E + A\mu +B\mu^2
\end{equation}
where $\mu=\eta-1$ and $\eta=\rho / \rho_0$ and $a$, $b$, $A$ and $B$ are Tillotson parameters.  We further assume  that $\mu$ remains small and therefore $\mu^2<<\mu$. This assumption is motivated by the fact that according to Eq.~(\ref{meos}) $\mu$ is given by $\mu=1-\rho_s/\rho_{s0}$ and even though the matrix density is not constant (as discussed above), the variation remains small as long $P<P_s$ (about 1\% in Fig.~\ref{fig:hugprho}). In fact, it can be shown that if $P_s << A$ then $(\rho_s - \rho_{s0}) / \rho_{s0} << 1$. According to Eq.~(\ref{meos}), the pressure is computed using the matrix density ($\mu=1-\rho_s/\rho_{s0}$), and therefore $\mu^2\simeq 0$ is a reasonable assumption.

Finally, we only want to consider cases where the energy $E$ remains small compared to the parameter $E_0$. The energy at $P=P_s$ can be estimated by: 
$$E = \frac{1}{2} P_s (\frac{1}{\rho_0}-\frac{1}{\rho}) \simeq \frac{1}{2} P_s (\frac{\alpha_0}{\rho_{s0}}-\frac{1}{\rho_{s0}}) = \frac{1}{2} P_s \frac{\alpha_0-1}{\rho_{s0}}$$ and therefore it is required that 
\begin{equation}
 E \simeq \frac{1}{2} P_s \frac{\alpha_0-1}{\rho_s0} << E_0.
\end{equation}
For $P_s$=7$\times$10$^9$ dyn/cm$^2$, $\rho_{s0}$=2.7 g/cm$^3$ and $\alpha_0$=3.0 we get $E\simeq$ 2.6$\times$10$^9$ erg/g which is indeed small compared to a typical value of  $E_0$ (e.g. 5$\times$10$^{10}$ erg/g for aluminium) and so this condition is fullfilled.

Using the above assumptions, we can rewrite Eq.~(\ref{eostpc}) in the very simple form
\begin{equation}
     P = c \rho E + A\mu
\end{equation}
where $c=a+b \simeq constant$. 

\section*{ACKNOWLEDGMENTS}
We are grateful to A.M. Nakamura and K. Hiraoka for providing us the crush-curve of pumice material that they 
measured in Kobe University. We also thank G. Collins and S. Sirono for their constructive reviews.
M.J. and W.B. gratefully acknowledge partial support from the Swiss National Science 
Foundation and from the Rectors' Conference of the Swiss Universities. M.J. acknowledges support from Kobe University 
(Japan) through the 21st Century COE (Center of Excellence) Program "Origin and Evolution of Planetary Systems".  
P.M. acknowledges support from the french Programme National de Plan\'etologie, from the Japanese Society 
for the Promotion of Science (JSPS) Invitation Fellowship for Research in Japan 2007, and with M.J. from the CNRS-JSPS cooperation program 2008-2009.

\section*{References}

\begin{description}

\item
A'Hearn, M.F., and Combi, M.R. 2007. Deep Impact at Comet Tempel 1. Icarus 191, 1-3.

\item{}
Benz, W., 1990. Smooth Particle Hydrodynamics - a review. In: J.R. Buchler (Ed.), 
The numerical modelling of nonlinear stellar pulsations: problems and prospects, p. 269.

\item{}
Benz, W., Asphaug, E., 1994. Impact simulations with fracture. I. Method and tests. Icarus 107, 98-116.

\item{}
Benz, W., Asphaug, E., 1995. Simulations of brittle solids using smooth particle hydrodynamics. Comput.
\ Phys.\ Comm. 87, 253-265.

\item{}
Benz, W., Jutzi, M., 2006. Collision and impact simulations including porosity. In: Milani, A., Valsecchi, G.B., 
Vokrouhlick\'y, D. (Eds.), Near-Earth Objects, Our Celestial Neighbors: Opportunity and Risk, IAU Symposium 236, 
in press. 

\item{}
Britt, D.T., Consolmagno, G.J., Merline, W.J., 2006. Small body density and porosity: new data, new insights. 
LPSC Abstract 2214.  

\item{}
Carroll, M.M., Holt, A.C., 1972. Suggested modification of the $P$ - $\alpha$ model for porous materials. 
J. Appl. Phys. 43, 759-761.

\item{}
Grady, D.E., Kipp, M.E., 1980. Continuum modelling of explosive 
fracture in oil shale. Int. J. Rock Mech. Min. Sci. \& 
Geomech. Abstr. 17, 147--157.

\item{}
Herrmann, W., 1969. Constitutive equation for the dynamic compaction of ductile porous materials. 
J. Appl. Phys., 40, 2490-2499.

\item{}
Housen, K.R., Holsapple, K.A., Voss, M.E., 1999. Compaction as the origin of the unusual craters on the asteroid 
Mathilde. Nature 402, 155-157.

\item{}
Housen, K.R., Holsapple, K.A., 2003. Impact cratering on porous asteroids. Icarus 163, 102-119. 

\item{}
Libersky, L.D., Petschek, A.G., 1991. Smooth Particle Hydrodynamics with strength of materials. In: Trease, 
Fritts, Crowley (Eds.), Proc. Next Free-Lagrange Method, Lecture Notes in Physics 395, Springer-Verlag, Berlin, 
pp. 248-257.

\item{}
Melosh, H.J., 1989. Impact cratering: a geological process. Oxford University Press, Oxford, UK. 

\item{}
Michel, P., Benz, W., Tanga, P., Richardson, D.C., 2001. Collisions
and gravitational reaccumulation: Forming asteroid families and
satellites.  Science 294, 1696-1700.

\item{}
Michel, P., Benz, W., Richardson, D.C., 2003. Disruption of
fragmented parent bodies as the origin of asteroid
families. Nature 421, 608-611.

\item{}
Monaghan, H.J., 1992. Smooth Particle Hydrodynamics. ARAA 30, 543-574.

\item{}
Richardson, D.C., Quinn, T., Stadel, J., Lake, G., 2000. Direct
large-scale $N$-body simulations of planetesimal
dynamics. Icarus 143, 45-59.

\item{}
Rickman, H., 1998. Composition and physical properties of comets. 
In: Solar System Ices, Astrophysics and space science library (AASL) Series 227, Kluwer Academinc Publishers, 
Dordrecht, p. 395.

\item{}
Sirono, S., 2004. Conditions for collisional growth of a grain aggregate. Icarus 167, 431-452

\item{}
Tillotson, J.H., 1962. Metallic equations of state for hypervelocity
impact. General Atomic Report GA-3216, July 1962.

\item{}
Weibull, W.A., 1939. A statistical theory of the strength of material
(transl.). Ingvetensk.\ Akad.\ Handl. 151, 5-45.

\item{}
Wuennemann, K., Collins, G.S., Melosh, H.J., 2006. A strain-based porosity model for use in hydrocode simulations 
of impacts
and implications for transient crater growth in porous targets. Icarus 180, 514-527.

\item{}
Wurm, G., Paraskov, G., Krauss, O., 2005. Growth of planetesimals by impacts at $\approx$ $25$ m$/$s. 
Icarus 178, 553-563.

\item{}
Yeomans, D.K., et al., 1997. Estimating the mass of asteroid 253 Mathilde from tracking data during the 
NEAR flyby. Science 278, 2106-2109.

\item{}
Zel'dovitch, Y.B., Raizer, Y.P., 1967. The physics of shock 
waves and high temperature hydrodynamic phenomena. Academic Press, 
New--York, USA.

\end{description}

\newpage
\begin{table}[!h]
\begin{center}
\begin{tabular}{|lcc|}
\hline
$P_e$ &\vline & 8e8 dyn/cm$^2$\\
$P_s$&\vline & 7e9 dyn/cm$^2$\\
$c_0$&\vline & 5.35e5 cm/s\\
$c_e$&\vline & 4.11e5 cm/s\\
\hline
\end{tabular}
\end{center}
 \caption{Parameters used in our porosity model for porous 
aluminium with an initial distention $\alpha_0=1.275$. Definitions of parameters are given in the text.}
\label{porosityalu}
\end{table}

\newpage

\newpage
\begin{table}[!h]
\begin{center}
\begin{tabular}{|lcc|}
\hline
$\alpha_0$&\vline & 4.64\\
$\alpha_t$&\vline & 1.90\\
$P_e$ &\vline & 1.00e7 dyn/cm$^2$\\
$P_t$&\vline & 6.80e9 dyn/cm$^2$\\
$P_s$&\vline & 2.13e9 dyn/cm$^2$\\
\hline
\end{tabular}
\end{center}
 \caption{Parameters used to fit the crush-curve of pumice.}
\label{porositypumice}
\end{table}

\newpage

\noindent
{\bf Figure 1:} Pressure versus  density (top) and pressure versus matrix-density 
(bottom) for a material with low ( $\alpha_0=1.275$) and high ($\alpha_0=3.0$) porosity. In the case of a high 
porosity, the density does never reach the initial matrix density (dashed horizontal line) and the 
matrix density itself decreases at a certain point, before $\alpha=1$ (dashed vertical line).

\medskip

\noindent
{\bf Figure 2:} Comparison of the pressure-distention relationship for different compaction models. The $P-alpha$ model assumes a quadratic function for $\alpha=\alpha(P)$. For the $\rho-alpha$ model we use a function $\alpha=\alpha(\rho)$ which follows from $\alpha=\alpha(P)$ by neglecting the thermal pressure. In the $\epsilon-alpha$ model we assume a constant matrix density ($\kappa=1$). In the low porosity case (top), full compaction is reached with all three models. In the high porosity case (bottom), only the $P-alpha$ model leads to full compaction. The minimal distention obtained using the constant matrix density assumption is about $\alpha$=1.5.

\medskip

\noindent
{\bf Figure 3:} Pressure-distention relationship for different values of $c\simeq a + b$ (Tillotson parameters) using the constant matrix density assumption. For low porosities, the influence of $c$ is rather small. In the high velocity case, the value of $c$ determines whether or not full compaction is possible at moderate pressures.

\medskip

\noindent
{\bf Figure 4:} Relation between distention and damage.

\medskip

\noindent
{\bf Figure 5:} SPH Simulation of a 1D compaction wave in a 3D cylinder composed of porous aluminium. There are two 
waves (top): an elastic precursor followed by the compaction (shock) wave. Bottom: the main decrease of the distention is due to the compaction wave. 

\medskip

\noindent
{\bf Figure 6:} Simulation of compression of a porous sample shown when $\alpha=\alpha_0=4.64$ (top) and 
during the compression (bottom) when $\alpha=2.2$. The sample is represented by the dark 
particles.

\medskip

\noindent
{\bf Figure 7:} Crush-curve of porous pumice measured by A.M. Nakamura and K. Hiraoka 
at Kobe University and obtained by a simulation using the $P-alpha$ and $\rho-alpha$ model.

\medskip

\noindent
{\bf Figure 8:} Simulation of a Deep Impact-like impact in porous ice with $\alpha_0=3.0$ (top) and non porous ice with the same initial density (bottom). Dark particles have a vertical velocity greater than 2 m/s.

\medskip

\noindent
{\bf Figure 9:} Cumulated mass of ejecta as a function of the ejection velocity as a result of a Deep Impact-like impact. The same initial density (i.e., the same mass) was used for the porous and non porous cases.

\medskip

\noindent
{\bf Figure 10:} Impact in a basalt target with an initial distention of $\alpha_0=3.0$ (black). Full compaction ($\alpha=1$, white) is only reached using 1 model (top). Using model 2 (middle) or 3 (bottom), there are no fully compacted particles. For the definition of the models, see text.

\medskip

\noindent
{\bf Figure 11:} Distention as a function of distance (negative z-direction) obtained by model 1/2/3.

\medskip

\noindent
{\bf Figure 12:} Cumulative volume of ejected material for the models 1/2/3. For comparison, the total volume which was compacted is also shown for the three cases (horizontal lines).

\medskip

\noindent
{\bf Figure 13:} Cumulative volume of ejected material for model 1 using $P_s$ = 2$\times$10$^9$ dyn/cm$^2$ and $P_s$ = 1$\times$10$^8$ dyn/cm$^2$. The horizontal lines indicate the total volume which was compacted in each case.

\newpage

\newpage

\newpage

\newpage

\newpage

\newpage

\newpage

\newpage

\newpage

\newpage

\newpage

\newpage

\newpage

\end{document}